\documentclass{jfm}
\usepackage{graphicx}
\usepackage{natbib}
\usepackage{amsbsy}
\usepackage{amsmath}
\usepackage{amssymb}
\usepackage{bm}
\usepackage[dvips]{color}

\title[Local boundary layer scales in turbulent Rayleigh-B\'{e}nard convection]{Local boundary layer scales in turbulent Rayleigh-B\'{e}nard convection}

\author[Janet D. Scheel and J\"org Schumacher]{Janet D. Scheel$^1$ and   J\"org Schumacher$^2$}
 \affiliation{$^1$Department of Physics, Occidental College, 1600 Campus Road, M21, Los Angeles, California 90041, USA \\[\affilskip]
                   $^2$Institut f\"ur Thermo- und Fluiddynamik, Postfach 100565, Technische Universit\"at Ilmenau, D-98684 Ilmenau, Germany}
\date{\today}
\begin{document}
\maketitle

\begin{abstract}

We compute {\em fully local boundary layer scales} in three-dimensional turbulent Rayleigh-B\'enard convection. These scales are directly connected to the highly intermittent fluctuations of the fluxes of 
momentum and heat at  the isothermal top and bottom walls and are statistically distributed around the corresponding mean thickness scales. 
The local boundary layer scales also reflect the strong spatial inhomogeneities of both boundary layers due to the large-scale, but complex and 
intermittent, circulation that builds up in closed convection cells. Similar to turbulent boundary layers, we define {\em inner scales} based on local shear 
stress which can be consistently extended to the classical viscous scales in bulk turbulence, e.g. the Kolmogorov scale, and {\em outer scales} based on 
slopes at the wall. We discuss the consequences of our generalization, in particular the scaling of our inner and outer boundary layer thicknesses and the 
resulting shear Reynolds number  with respect to Rayleigh number. The mean outer thickness scale  for the temperature field is close to the 
standard definition of a thermal boundary layer thickness. In the case of the velocity field, under certain conditions the outer scale 
follows a similar scaling as the Prandtl-Blasius type definition with respect to Rayleigh number, but differs quantitatively. The friction coefficient 
$c_{\epsilon}$ scaling is found to fall  right between the laminar and turbulent limits which indicates that the boundary layer exhibits transitional behavior.  Additionally, we conduct an analysis of the recently suggested dissipation layer thickness scales versus Rayleigh  
number and find a transition in the scaling.  All our investigations are based on highly accurate spectral element simulations which reproduce gradients and 
their fluctuations reliably. The study is done for a Prandtl number of $Pr=0.7$ and for Rayleigh numbers which extend over nearly five orders of magnitude,  
$3\times 10^5\le Ra \le 10^{10}$ in cells of  aspect ratio of one. We also performed one study of aspect ratio equal to three in the case of $Ra=10^8$. For 
both aspect ratios, we find that the scale distributions depend on the position at the plates where the analysis is conducted. 
\end{abstract}

\section{\label{sec:level1} Introduction}
The key to a deeper understanding of the mechanisms of transport of heat and momentum in turbulent Rayleigh-B\'{e}nard 
convection lies in a better access to the dynamics in the tiny boundary layers of the temperature and velocity fields (\citeauthor{Siggia1994}
\citeyear{Siggia1994}, \citeauthor{Ahlers2009} \citeyear{Ahlers2009}, \citeauthor{Chilla2012} \citeyear{Chilla2012}). The boundary layers (BL)
form in the vicinity of the isothermally heated plate at the bottom and the isothermally cooled plate at the top. With increasing Rayleigh 
number $Ra$ and thus with increasing thermal driving of Rayleigh-B\'{e}nard convection,  the BL thickness decreases similar to many (isothermal) wall-bounded 
flows where the BL thickness shrinks with increasing Reynolds number (see e.g. \citeauthor{Pope2000} \citeyear{Pope2000}).          

The full access to the three-dimensional structure and near-wall dynamics of the BL is still impossible for high-Rayleigh number
convection experiments. However in the last few years, successful steps into this direction have been made by monitoring the dynamical evolution in the
boundary layer in two-dimensional observation windows using high-resolution particle image velocimetry, such as in  \cite{Zhouetal2010}, \cite{Zhou2010}, and
\cite{duPuits2014}. Local heat flux measurements on small plate segments such as those by \cite{duPuits2010} or \cite{Kaiser2014} provide
an access to the fluctuating temperature gradient at the walls. Time-averaged local boundary layer profiles were measured by \cite{Lui1998} for 
$2\times 10^8<Ra<2\times 10^{10}$. But for Rayleigh numbers $Ra \gtrsim 10^{12}$, a direct experimental access to 
the boundary layer is not yet possible (\citeauthor{He2012} \citeyear{He2012}, \citeauthor{Urban2012} \citeyear{Urban2012}). Only the mean thermal 
boundary layer thickness can be reproduced from global heat flux measurements. Direct numerical simulations lack the access to this Rayleigh number 
regime when the aspect ratio of the cell remains larger than or equal to unity (\citeauthor{Hartlep2005} \citeyear{Hartlep2005}, \citeauthor{Bailon2010} 
\citeyear{Bailon2010}). Nevertheless, particularly in the last years several direct numerical simulation studies provided new insights into the structure of the boundary 
layers of both fields, e.g. in \cite{Reeuwijk2008b}, \cite{Zhouetal2010}, \cite{Stevens2010}, \cite{Stevens2012}, \cite{Wagner2012}, \cite{Scheel2012} and \cite{Shi2012}.  
The main outcome is that the boundary layers obey a complex three-dimensional dynamics which is strongly time-dependent and coupled to the large-scale circulation 
in the convection system as well as the small-scale fluctuations in the bulk and the emission of thermal plumes.    

The classical definition of a boundary layer thickness is traditionally defined (\cite{Ahlers2009,Shi2012}) as the intersection point between the tangent to 
the time-averaged profile at the plates and the first local maximum of  the velocity profile (for $\delta_v$) or the bulk value of the temperature profile (for $\delta_T$). 
This is known as the slope method. For alternate definitions see \cite{Ahlers2009} or \cite{Li2012}. Note that all standard definitions of boundary layer thickness, such 
as the displacement thickness,  assume the existence of a well-defined mean flow. The thickness of both boundary layers are central quantities in mean field theories 
for the global heat and momentum transfer (see e.g. 
\citeauthor{Grossmann2000} \citeyear{Grossmann2000}). They are required to divide the dissipation of kinetic energy and thermal variance 
into bulk and boundary layer dominated volume fractions. The mean thermal boundary layer thickness scales inversely with the Nusselt number
$Nu$, the dimensionless measure of the global heat transfer and is given by 
\begin{equation}
\delta_T=\frac{H}{2 Nu}\,,
\label{TBL}
\end{equation}
with $H$ being the height of the convection layer (or convection cell). The mean boundary layer thickness  of the velocity field scales with the square root of the Reynolds number $Re$ and is given by
\begin{equation}
\delta_v=\frac{aH}{\sqrt{Re}}\,,
\label{vBL}
\end{equation}
where $a$ is a free parameter which is adjusted to the case of a Prandtl-Blausius type boundary layer (\citeauthor{Prandtl1905} \citeyear{Prandtl1905}, \citeauthor{Blasius1908}
\citeyear{Blasius1908}). There are however applications in nature where these concepts are not applicable, particularly since the derivation relies on no-slip
boundary conditions at the isothermal plates at the top and bottom. One such example is the dynamics of the mantle of the Earth where convection
models use stress-free boundaries without momentum flux (see e.g. \citeauthor{Trompert1998} \citeyear{Trompert1998}) and for which definition 
(\ref{vBL}) does not work. This was one motivation of \cite{Petschel2013}
  to generalize the definition of a boundary 
layer thickness to that of a {\em dissipation layer thickness} (DL). The thickness of the dissipation layer is determined by
\begin{equation}
\label{disslayer}
\langle\epsilon(z=d_{v})\rangle_{A,t} = \langle\epsilon\rangle_{V,t}\ \ {\rm{and}}\ \ \langle\epsilon_T(z=d_{T})\rangle_{A,t} = \langle\epsilon_T\rangle_{V,t}.
\end{equation}
Then the smallest distance from a boundary plate to the intersection of these  equations defines the DL scales $d_{v}$ 
and $d_{T}$. Here, $\epsilon$ is the rate of kinetic energy dissipation and $\epsilon_T$ the thermal dissipation rate (see below for exact definitions).
Notations $\langle\cdot\rangle_{A,t}$ and $\langle\cdot\rangle_{V,t}$ denote averages over horizontal planes in combination with time and volume in
combination with time, respectively. This definition requires the measurement of profiles of both dissipation rates which is not applicable in experiments.  
 The two methods to calculate thicknesses of boundary layers, either via Eqns. (\ref{TBL}) and (\ref{vBL}) or via (\ref{disslayer}), result in  mean thickness scales. They do not incorporate the strong spatial inhomogeneities across the isothermal plates.
Furthermore, it is well-known from experiments (\cite{Lui1998}, \cite{Qiu1998}, \cite{Zhouetal2010}, \citeauthor{Zhou2010} \citeyear{Zhou2010}) and simulations (\cite{Zhouetal2011}, \cite{Wagner2012}, \cite{Stevens2012}, \citeauthor{Shi2012} \citeyear{Shi2012})
that these thicknesses vary strongly in time and that this is caused by the local detachment of thermal plumes as well as fluctuations in the large-scale circulation and in the bulk.

In this work, we use therefore a {\em fully local boundary layer scale} definition which incorporates these spatial inhomogeneities. These local boundary  layer scales are distributed around a mean scale and thus reflect the strong spatial intermittency  of the gradients in the vicinity of the walls in the turbulent convection flow. Our approach is  fully local since it is based on gradients evaluated at the plate with spectral accuracy. We define two classes of local boundary layer scales, {\em inner} and {\em outer} scales. Our  definition for the outer boundary layer scale for the velocity and temperature fields is based on the fluxes of momentum and heat at the wall, respectively.  These BL scales are thus length scales which build on an inverse gradient for both fields (excluding zero magnitudes). In addition these local methods can be applied even in the absence of a well-defined mean flow, which is the case in turbulent thermal convection. In the case of the temperature field, the mean of this outer scale 
 is close to Eq. (\ref{TBL}) which relates the thermal boundary layer thickness to the Nusselt number. In the case of the velocity field, our mean outer scale definition 
avoids the use of the coefficient $a$ in combination with the prescribed functional dependence on the Reynolds number (see Eq. (\ref{vBL}.)) This will cause a 
scaling of the mean velocity boundary layer thickness that is qualitatively similar to that of Eq. (\ref{vBL}), but differs quantitatively.  

The definitions of the inner scales can be consistently  related to the local dissipation and diffusion scales which 
have been developed and investigated for bulk turbulence in several systems (\citeauthor{Schumacher2005} \citeyear{Schumacher2005},
\citeauthor{Schumacher2007} \citeyear{Schumacher2007}, \citeauthor{Zhou2010a} \citeyear{Zhou2010a}, \citeauthor{Hamlington2012} \citeyear{Hamlington2012},
\citeauthor{Scheel2013} \citeyear{Scheel2013}). Clearly, velocity gradients in the form of the wall-shear stress enter these definitions again, but now related to molecular viscosity and 
thermal diffusivity, respectively. Finally, we will compare our analysis with the dissipation layer approach. While in \cite{Petschel2013} trends of the dissipation
layer thickness with respect to Prandtl number at fixed Rayleigh number are discussed, we investigate the dependence on the Rayleigh number for a 
Prandtl number of convection in air.        

The outline of the manuscript is as follows. In the next section, we will discuss the equations of motion, list essential definitions and describe the numerical method
in brief. Results of the mean global transport of heat and momentum are also listed. The third section summarizes the definitions of the local boundary layer 
scales and relates them to classical equations. The fourth section discusses our results. We present the distributions
of the scales and compare the means with classical thickness equations, particularly with respect to the scaling with Rayleigh number. We also compute the resulting shear Reynolds number and its scaling with Rayleigh number. Furthermore, the spatial averaging is conducted partly locally. Therefore we will analyze scale distributions in different subvolumes, including in an aspect-ratio-of-three cell and compare the findings to a cylindrical cell with aspect ratio one. This analysis is followed by
a dissipation layer analysis which we conduct here for varying Rayleigh number. This analysis is followed by a final study of the friction coefficient and compared with 
existing data from \cite{Verzicco2003} and \cite{Wei2013}. We conclude with a final summary and a brief discussion.

\section{\label{sec:level2} Methods}
We solve the three-dimensional equations of motion in the Boussinesq approximation numerically. The dimensionless form of the equations\footnote{Note that all scaled variables will be noted by tildes in this  section.} is based on the
height of the cell $H$, the free-fall velocity $U_f=\sqrt{g \alpha \Delta T H}$ and the imposed temperature difference $\Delta T$. The three control parameters 
of Rayleigh-B\'enard convection are the Rayleigh number $Ra$, the Prandtl number $Pr$ and the aspect ratio $\Gamma=D/H$ with the diameter $D$. The resulting equations are:
\begin{eqnarray}
\label{ceq}
\tilde {\bm \nabla}\cdot\tilde {\bf u}&=&0\,,\\
\label{nseq}
\frac{\partial\tilde{\bf u}}{\partial \tilde t}+(\tilde{\bf u}\cdot\tilde{\bm\nabla})\tilde{\bf u}
&=&-\tilde{\bm \nabla} \tilde p+\sqrt{\frac{Pr}{Ra}} \tilde{\bm \nabla}^2\tilde{\bf u}+ \tilde T {\bf e}_z\,,\\
\frac{\partial \tilde T}{\partial \tilde t}+(\tilde {\bf u}\cdot\tilde {\bm \nabla}) \tilde T
&=&\frac{1}{\sqrt{Ra Pr}} \tilde{\bm \nabla}^2 \tilde T\,,
\label{pseq}
\end{eqnarray}
where
\begin{equation}
Ra=\frac{g\alpha\Delta T H^3}{\nu\kappa}\,,\;\;\;\;\;\;\;\;Pr=\frac{\nu}{\kappa}\,.
\end{equation}
The variable $g$ stands for the  acceleration due to gravity, $\alpha$ is thermal expansion coefficient, $\nu$ is the kinematic viscosity, and 
$\kappa$ is thermal diffusivity. We use aspect ratios of  $\Gamma=1$ and 3. At all walls no-slip boundary conditions for the 
fluid are applied, i.e., ${\bf u}=0$. The side walls are thermally insulated, i.e., the normal derivative of the temperature field vanishes,
$\partial T/\partial {\bf n}=0$. The top and bottom plates are held at constant dimensionless temperatures $\tilde T=0$ and 1, respectively. In response to
the input parameters $Ra$, $Pr$ and $\Gamma$, a  heat flux is established from the bottom to the top plate.  
It is determined by the Nusselt number which is defined as  
\begin{equation}
Nu(\tilde z)=\sqrt{Ra Pr}\, \langle\tilde u_z \tilde T\rangle_{A,t}-\frac{\partial\langle \tilde T\rangle_{A,t}}{\partial \tilde z}\,.
\label{Nusselt}
\end{equation}
The vertical average of $Nu(\tilde z)$ results in the volume averaged Nusselt number $Nu_V$:
\begin{equation}
Nu_V=1+\sqrt{Ra Pr}\langle \tilde u_z \tilde T\rangle_{V,t}\,.
\label{Nusselt2}
\end{equation}
The value $Nu_V$ has to be equal to $Nu(\tilde z)$  for all $\tilde z\in [0,1]$. The momentum transport is expressed by the Reynolds number which is defined as
\begin{equation}
Re=\sqrt{\frac{Ra}{Pr}\,\langle\tilde{\bf u}^2\rangle_{V,t}}\,.
\label{Reynolds}
\end{equation}
The thermal dissipation rate is given by  
\begin{equation}
\tilde{\epsilon}_T(\tilde{\bf x},\tilde t) = \frac{1}{\sqrt{Ra Pr}}\left(\frac{\partial \tilde T}{\partial \tilde x_j}\right)^2\,,
\label{thermal1}
\end{equation}
and the kinetic energy dissipation rate is defined as 
\begin{equation}
\tilde\epsilon(\tilde{\bf x}, \tilde t) =\frac{1}{2} \sqrt{\frac{Pr}{Ra}}\left(\frac{\partial \tilde u_i}{\partial \tilde x_j}+\frac{\partial \tilde u_j}{\partial \tilde x_i}\right)^2=2 \sqrt{\frac{Pr}{Ra}} \tilde{S}_{ij}\tilde{S}_{ji}\,.
\label{kinetic1}
\end{equation}
where $\tilde{S}_{ij}$ is the rate of strain tensor.

The equations are numerically solved by the Nek5000 spectral element method package which has been adapted to our 
problem. The code employs second order time-stepping, using a second-order backward difference formula. The whole set of
equations is transformed into a weak formulation and discretized with the particular choice of spectral basis functions (\citeauthor{Fischer1997} 
\citeyear{Fischer1997}, \citeauthor{Deville2002} \citeyear{Deville2002}). The resulting linear, symmetric Stokes problem is solved implicitly. 
This system is split, decoupling the viscous and pressure steps into independent symmetric positive definite subproblems which are solved 
either by Jacobi (viscous) or multilevel Schwartz (pressure) preconditioned conjugate gradient iteration. Fast parallel solvers based on 
direct projection or more scalable algebraic multigrid are used for the coarse-grid solve that is part of the pressure preconditioner. All derivatives are calculated spectrally  when computing the local boundary layer thicknesses in Eqns. (\ref{tempBL1}), (\ref{velBL1}) and (\ref{velBL2}).
For further numerical details and comprehensive tests of the sufficient spectral resolution, we refer to \cite{Scheel2013}. 

\begin{table}
\begin{center}
\begin{tabular}{cccccccc}
 $Ra$ & $\Gamma$ & $N_s$  &  $N_e$  &  $N$  & $Nu_V\pm \Delta Nu_v$ & $Re\pm \Delta Re$ & $\tilde{u}_{rms}\pm \Delta\tilde{u}_{rms}$\\
\hline
$3\times 10^5$*  &1   & 401 &  3,520 &  11& 5.80 $\pm 0.03$ & 116 $\pm 1$  & 0.177 $\pm 0.001$\\
$5\times 10^5$*  &1   &  401 & 3,520 &  11&  6.90 $\pm 0.13$ & 151 $\pm 2$ & 0.179 $\pm 0.002$\\
$7\times 10^5$*  &1   &  407 & 3,520 &  11&  7.78 $\pm 0.05$ & 179 $\pm 1$ & 0.180 $\pm 0.002$\\
$1\times 10^6$  &1   &  300 & 30,720 &  7& 8.65 $\pm 0.06$ & 214 $\pm 6$ & 0.179 $\pm 0.004$\\
$5\times 10^6$  &1   &  340 &  30,720 & 7 & 13.79 $\pm 0.17$ & 483 $\pm 1$ & 0.181 $\pm 0.001$\\
$1\times 10^7$  &1   &  230 &  30,720 & 11 & 16.77 $\pm 0.01$& 675 $\pm 3$  & 0.179 $\pm 0.001$\\
$5\times 10^7$  &1   &  192 &  30,720 & 13 & 25.8 $\pm 0.3$ & 1490 $\pm 40$ & 0.176 $\pm 0.004$\\
$1\times 10^8$  &1   &  87   & 256,000 & 11& 31.4 $\pm 1.3$ & 2070 $\pm 60$ & 0.173 $\pm 0.005$\\
$1\times 10^8$  &3   &  62    & 2,304,000 & 9 & 31.1 $\pm 0.6$& 2310 $\pm 30 $ & 0.194 $\pm 0.002$\\
$1\times 10^9$  &1   &  92   & 875,520 & 11& 63 $\pm 4$ & 6240 $\pm 140$ & 0.165 $\pm 0.004$\\
$1\times 10^{10}$* &1   &  41   & 2,374,400 & 11 & 127 $\pm 6$& 19300 $\pm 900$ & 0.161 $\pm 0.007$\\
\hline
\end{tabular}
\end{center}
\caption{A summary of the parameters used for the convection runs.
We list the Rayleigh number $Ra$, the aspect ratio $\Gamma$, the number of statistically 
independent snapshots $N_s$, the number of spectral elements $N_e$, polynomial order $N$ in each space direction and for each element, the 
Nusselt number $Nu_V$ (see Eq. (\ref{Nusselt2})),  the Reynolds number $Re$ (see Eq. (\ref{Reynolds})) and the root mean square velocity
obtained in the whole cell volume $V$, i.e. $\tilde{u}_{rms}=\sqrt{\langle \tilde{\bf u}^2\rangle_{V,t}}$.  All runs are conducted at $Pr=0.7$. Stars indicate simulations which are new (the rest were first presented in \cite{Scheel2013}). The error 
bars in the last three columns have been obtained by evaluating the results over the first and second halves of the corresponding data set separately and taking
the difference of both results subsequently.}
\label{Tab1}
\end{table}

In table \ref{Tab1} we summarize the main parameters of the simulation runs. The total number of mesh cells is calculated by $N_e N^3$ and becomes 
larger than $4\times 10^9$ for the largest runs. We detect the following power laws for the global transport of heat and momentum. The Reynolds number
follows $Re=(0.25\pm 0.01)\times Ra^{0.49\pm 0.01}$, the Nusselt number yields $Nu=(0.15 \pm 0.01)\times Ra^{0.29\pm 0.01}$. Compared to \cite{Scheel2013}, we extended the series of DNS
runs by additional data points at the lower and higher Rayleigh numbers ($3-7\times 10^5, 5\times 10^7$, and $1 \times 10^{10}$ are new)
and have run a few time series longer to improve statistics. We also list 
the root mean square velocity which is obtained as a combined volume-time average in the whole cell volume $V$, i.e. $\tilde{u}_{rms}=\sqrt{\langle \tilde{\bf u}^2\rangle_{V,t}}$. 

\section{\label{sec:level3} Local boundary layer scales}
\subsection{Outer local boundary layer scales}
The following definitions are based on the current densities of heat and momentum at the wall, respectively. The heat current density at the plate 
is purely diffusive and given by 
\begin{equation}
\label{flux1}
J^{\rm{heat}}(x, y)=-\kappa \frac{\partial T}{\partial z}\Bigg|_{z=0}\,.
\end{equation}
The local temperature slope scale is associated with the inverse (nonzero) gradient and can be defined as
\begin{equation}
\label{tempBL}
\lambda^{o}_T(x,y) =\dfrac{\kappa\dfrac{\Delta T}{2}}{|J^{\rm{heat}}(x, y)|} = \frac{H}{2} \,\Bigg| \frac{\partial \tilde{T}}{\partial \tilde{z}}\Bigg|_{\tilde z=0}^{-1}\,.
\end{equation}
The superscript ``{o}" stands for {\em outer local boundary layer scale}.
The factor $\frac{1}{2}$ is included since the thickness at each of the two isothermal plates is  related to one half of the total temperature jump. 

Eq. (\ref{tempBL}) is thus a local boundary layer (thickness) scale based on the local slope at each grid point on the plate. It is equivalent to the slope method for an 
instantaneous local profile, but is much easier to calculate numerically and it is more precise since derivatives are calculated spectrally.

The mean value of this temperature slope scale follows as
\begin{equation}
\label{tempBL1}
\langle\tilde\lambda^{o}_T\rangle=\int_0^{\infty}\,\tilde\lambda^{o}_T\,p(\tilde\lambda^{o}_T) \,\mbox{d}\tilde\lambda^{o}_T= 
\frac{1}{2}\Bigg\langle\Bigg|\frac{\partial \tilde{T}}{\partial \tilde{z}}\Bigg|_{\tilde z=0}^{-1}\Bigg\rangle\,,  
\end{equation}
where $p(\tilde\lambda^{o}_T)$ is the probability density function (PDF) of $\lambda^{o}_T(x,y)$. 
It remains to be verified in the simulations if the obtained mean scale coincides with the classical equation of the thermal boundary layer thickness, i.e., 
$\langle\tilde\lambda^{o}_T\rangle\approx 1/(2 Nu)=\tilde{\delta}_T$. Note that it does not follow rigorously (see also Eq.(\ref{Nusselt})). 

\begin{figure}
\centering
\includegraphics[width=0.95\textwidth]{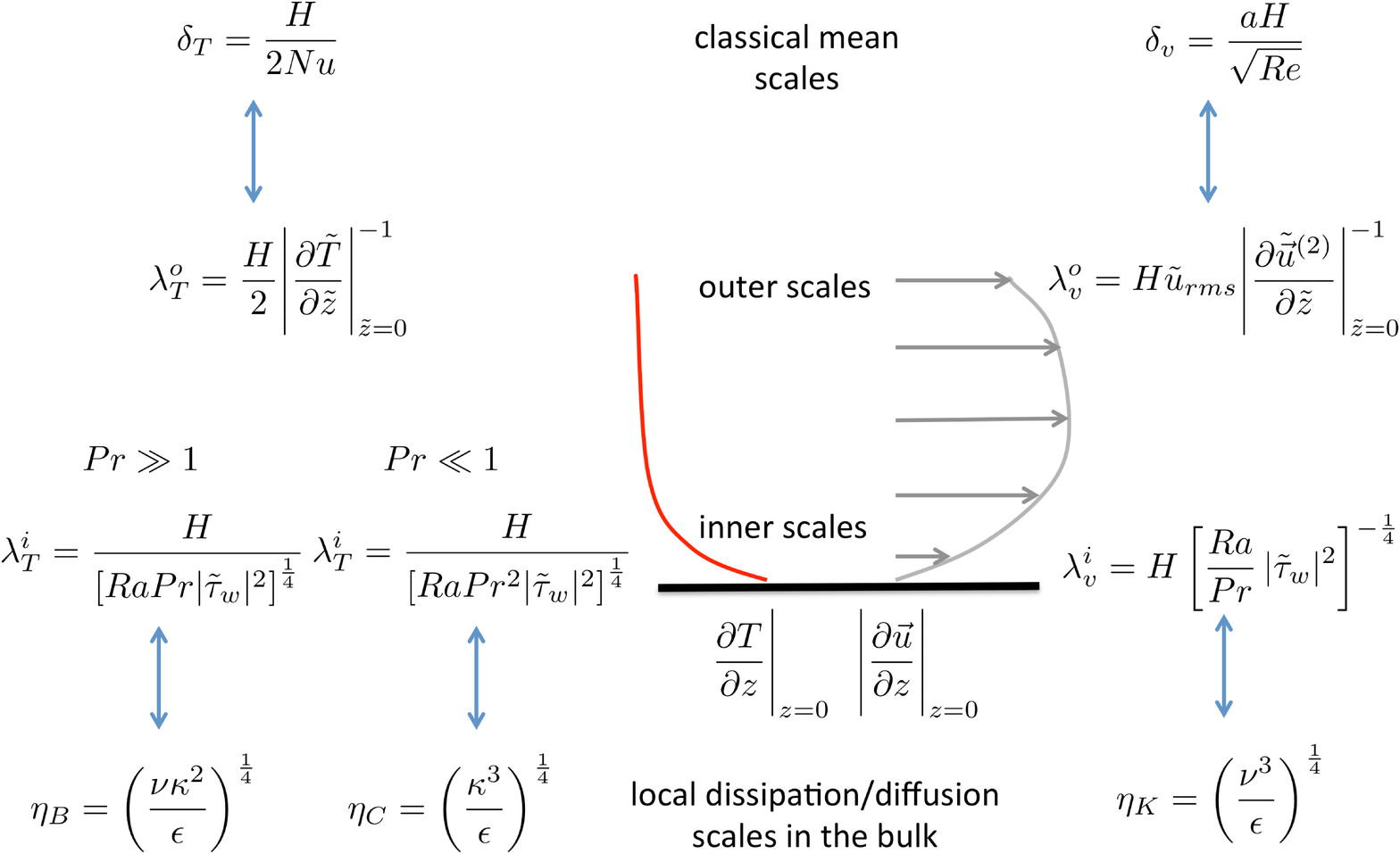}
\caption{A summary diagram of the various connections between the definitions of local boundary layer scales. Similar to turbulent boundary 
layer theory we suggest outer and inner local scales which can be related to classical mean thickness scales and local dissipation scales 
in the bulk, respectively. The vertical double-headed arrows indicate the direct connections to other definitions.}
\label{fig1}
\end{figure}

We proceed in 
the same way for the velocity field. The momentum current density at the wall has two non-vanishing components and the two-dimensional vector field is given by
\begin{equation}
\label{flux2}
{\bf J}^{\rm{mom}}(x, y)= \nu \frac{\partial {\bf u}^{(2)}}{\partial z}\Bigg|_{z=0}\,,
\end{equation}
with ${\bf u}^{(2)}=(u_x, u_y)$. The magnitude of ${\bf J}^{\rm{mom}}$ is  the (kinematic) wall shear stress $\tau_w$ when an average over plane and time is taken. 
Two contributions to the local wall shear stress remain in the case of no-slip 
boundary conditions. In analogy to definition (\ref{tempBL}), we define the local velocity boundary scale
\begin{equation}
\label{velBL}
\lambda^{o}_v( x, y) = \frac{\nu u_{rms}}{|{\bf J}^{\rm{mom}}(x, y)|}=H \tilde{u}_{rms}\Bigg | \frac{\partial \tilde{\bf u}^{(2)}}{\partial \tilde z}\Bigg |^{-1}_{\tilde z =0}\,,
\end{equation}
and obtain the following mean thickness
\begin{equation}
\label{velBL1}
\langle \tilde\lambda^{o}_v\rangle = \int_0^{\infty}\,\tilde\lambda^{o}_v\,p(\tilde\lambda^{o}_v) \,\mbox{d}\tilde\lambda^{o}_v =
\tilde{u}_{rms}\Bigg\langle\Bigg |\frac{\partial \tilde{\bf u}^{(2)}}{\partial \tilde z}\Bigg |^{-1}_{\tilde z =0}\Bigg\rangle\,.
\end{equation}
Eq. (\ref{velBL1}) is again equivalent to the slope method for an instantaneous local profile. However, this definition moves away from the classical thickness  equation which incorporates the well-known Reynolds-number dependence of laminar boundary layer 
theory. Nevertheless we have to introduce a scale factor. In contrast to the temperature 
scale definition (\ref{tempBL}), where the prefactor of $\frac{1}{2}$ is prescribed by the physical picture of two symmetric boundary layers for the Boussinesq case across 
which the whole sustained temperature difference drops, it is ab initio undetermined for the velocity. Our prefactor $\tilde u_{rms}$ takes therefore the role of the $a$ in the classical 
 equation  (\ref{vBL}) and as this $a$ needs to be determined from experiment (see e.g. \cite{Ahlers2009}), our approach will require the evaluation of $\tilde u_{rms}$ from 
simulation. There is a simple geometric picture behind this prefactor. Without any scale factor, we would get a slope scale for the velocity profile which is related to the free-fall velocity
$U_f$. As it has been also discussed in the caption of Table 1 of \cite{Bailon2010}, $U_f$ is more than a factor of 5 bigger than $u_{rms}$. As a consequence having no scale factor would 
overestimate our thickness scale. Hence we scale with $\tilde u_{rms}$. We will investigate the relation of our mean scale to the classical Prandtl-Blasius-type equation (\ref{vBL}) 
and will test if $\langle \tilde\lambda^{o}_v\rangle \approx a/\sqrt{Re}
=\tilde\delta_v$ holds. This discussion  will follow in subsection \ref{Scalinglaws} (see also Tab. \ref{Tab1}). 
The diagram in Fig. \ref{fig1} displays the connections between the classical thickness scales and our definitions for temperature and velocity.  

\subsection{Inner local boundary scales}
We proceed with the definitions of {\em inner local boundary scales}. The inner local velocity boundary scale is defined as follows
\begin{equation}
\label{velBL2}
\lambda^{i}_v( x, y) = \frac{\nu}{\sqrt{|{\tau}_w(x, y)|}}=\frac{H}{\sqrt{Re_f}}\left[ \left( \frac{\partial \tilde u_x}{\partial \tilde z}\right)^2+\left(\frac{\partial \tilde u_y}{\partial \tilde z} \right)^2\right]^{-\frac{1}{4}}_{\tilde z =0}\,,
\end{equation}
where $Re_f=\sqrt{Ra/Pr}$ is a Reynolds number based on cell height $H$ and free-fall velocity $U_f$. The definition corresponds to the 
well-known inner scale from turbulent boundary layer theory, $\lambda_+=\nu/u_{\tau}$ with the (mean) friction velocity  $u_{\tau}$  (see e.g. 
\citeauthor{Pope2000} \citeyear{Pope2000}). One obtains the following mean thickness
\begin{equation}
\label{velBL2a}
\langle \tilde\lambda^{i}_v\rangle = \int_0^{\infty}\,\tilde\lambda^{i}_v\,p(\tilde\lambda^{i}_v) \,\mbox{d}\tilde\lambda^{i}_v =
\left(\frac{Pr}{Ra}\right)^{\frac{1}{4}}\langle |\tilde{\tau}_w|^{-\frac{1}{2}}\rangle\,.
\end{equation}
Here and in Fig. \ref{fig1}, we abbreviated $|\tilde{\tau}_w|=((\partial_{\tilde z} \tilde u_x)^2+(\partial_{\tilde z} \tilde u_y)^2)_{\tilde{z}=0}^{1/2}$.
The inner local velocity boundary layer scale is related to the local dissipation scale which has been defined in \cite{Schumacher2005} or \cite{Scheel2013}. 
In this framework, the (mean) Kolmogorov scale $\eta_K=\nu^{3/4}/\langle\epsilon\rangle^{1/4}$ was  generalized to a local dissipation scale
\begin{equation}
\label{velBL3}
\tilde\eta_K(\tilde{\bf x},\tilde{t}) = \left(\frac{Pr}{Ra}\right)^{\frac{3}{8}}\,\tilde{\epsilon}(\tilde{\bf x},\tilde{t})^{-\frac{1}{4}}\,.
\end{equation}
When applying definition (\ref{kinetic1}) and defining a local shear rate on the basis of the rate of strain tensor 
\begin{equation}
\label{norm}
\tilde S(\tilde{\bf x})=\sqrt{\tilde{S}_{ij}\tilde{S}_{ji}}\,,
\end{equation}
one ends up with  a consistent translation into our definition of the local inner velocity boundary thickness $\tilde\lambda^{i}_v(\tilde x,\tilde y)$
\begin{eqnarray}
\label{velBL3a}
\lim_{\tilde{z}\to 0}\tilde\eta_K(\tilde{\bf x},\tilde{t}) &=& \left(\frac{Pr}{Ra}\right)^{\frac{1}{4}}\,\lim_{\tilde{z}\to 0}\left[\sqrt{2}\,\tilde{S}(\tilde{\bf x},\tilde{t})\right]^{-\frac{1}{2}}\nonumber\\
&=& \left(\frac{Pr}{Ra}\right)^{\frac{1}{4}} \left[ \frac{1}{2}\left( \frac{\partial \tilde u_x}{\partial \tilde z}\right)^2+\frac{1}{2}\left(\frac{\partial \tilde u_y}{\partial \tilde z} \right)^2\right]_{z=0}^{-\frac{1}{4}} =\sqrt[4]{2} 
\,\tilde{\lambda}^{i}_v(\tilde x,\tilde y)\,.
\end{eqnarray}
Since the fourth root of 2 is close to unity, both scales practically coincide when approaching the boundary from the bulk.
Thus, the analysis which has been formerly conducted in the bulk of the convection cell (see \cite{Scheel2013}) can be systematically 
continued to the walls at the heating and cooling plates.

In the case of the temperature field, we will have to distinguish between small and  large Prandtl numbers. 
The inner scale of temperature field can then be obtained in a similar way to how the Corrsin and Batchelor diffusion scales are obtained 
in the regimes of low- and high-Prandtl-number convection, respectively (see e.g. \citeauthor{Groetzbach1983} \citeyear{Groetzbach1983},
\citeauthor{Schumacher2005} \citeyear{Schumacher2005}).   The (mean) Corrsin scale is given by $\eta_C = \kappa^{3/4}/\langle\epsilon\rangle^{1/4}
 = \eta_K/Pr^{3/4}$ and the (mean) Batchelor scale $\eta_B=\kappa^{1/2}\nu^{1/4}/\langle\epsilon\rangle^{1/4} = \eta_K/\sqrt{Pr}$, respectively.  Consequently, 
 we can define the following local scales
\begin{equation}
\label{TBLi1}
\tilde{\lambda}^{i}_T( \tilde x, \tilde y) =\frac{\tilde{\lambda}_v^{i}(\tilde{x},\tilde{y})}{\sqrt{Pr}}=(RaPr)^{-\frac{1}{4}}\left[ \left( \frac{\partial \tilde u_x}{\partial \tilde z}\right)^2+\left(\frac{\partial \tilde u_y}{\partial \tilde z} \right)^2\right]^{-\frac{1}{4}}_{\tilde z =0}
\;\;\;\mbox{for}\;\;\;Pr\ge 1\,,
\end{equation}
and
\begin{equation}
\label{TBLi2}
\tilde{\lambda}^{i}_T( \tilde x, \tilde y) =\frac{\tilde{\lambda}_v^{i}(\tilde{x},\tilde{y})}{Pr^{\frac{3}{4}}}=(RaPr^2)^{-\frac{1}{4}}\left[ \left( \frac{\partial \tilde u_x}{\partial \tilde z}\right)^2+\left(\frac{\partial \tilde u_y}{\partial \tilde z} \right)^2\right]^{-\frac{1}{4}}_{\tilde z =0}
\;\;\;\mbox{for}\;\;\;Pr\le 1\,.
\end{equation}
Since the inner boundary scales for temperature differ only by powers of $Ra$ and $Pr$ compared to $\lambda_v^{i}$, the distributions will just be shifted with respect of those of $\lambda_v^{i}$ (see e.g. \cite{Schumacher2005}). Therefore we will only calculate the inner local velocity scale $\lambda_v^{i}$ in this paper.

We refer once more to Fig. \ref{fig1} where our definitions are summarized. All the local scale definitions require us to exclude the zero-gradient points. In the case
of the temperature field,  zero gradients at the walls are excluded a priori due to solely diffusive transport. We have verified this in our data sets. In the case of 
the velocity boundary layer zero-gradient events remain at and below 1\% of all events. The number is found to decrease with increasing Rayleigh number.  
These local events are avoided only when averages over an ensemble or over an area in combination with time are taken as in the definitions
of mean thickness scales. We will come back to this point at the end of subsection \ref{cepschapter}. 

From now on, we will use the dimensionless quantities 
only and drop the tildes in all expressions for convenience.

\section{\label{sec:level4} Results}
\subsection{Distribution of local boundary layer scales across the boundary plates}
\begin{figure}
\centering
\includegraphics[width=\textwidth]{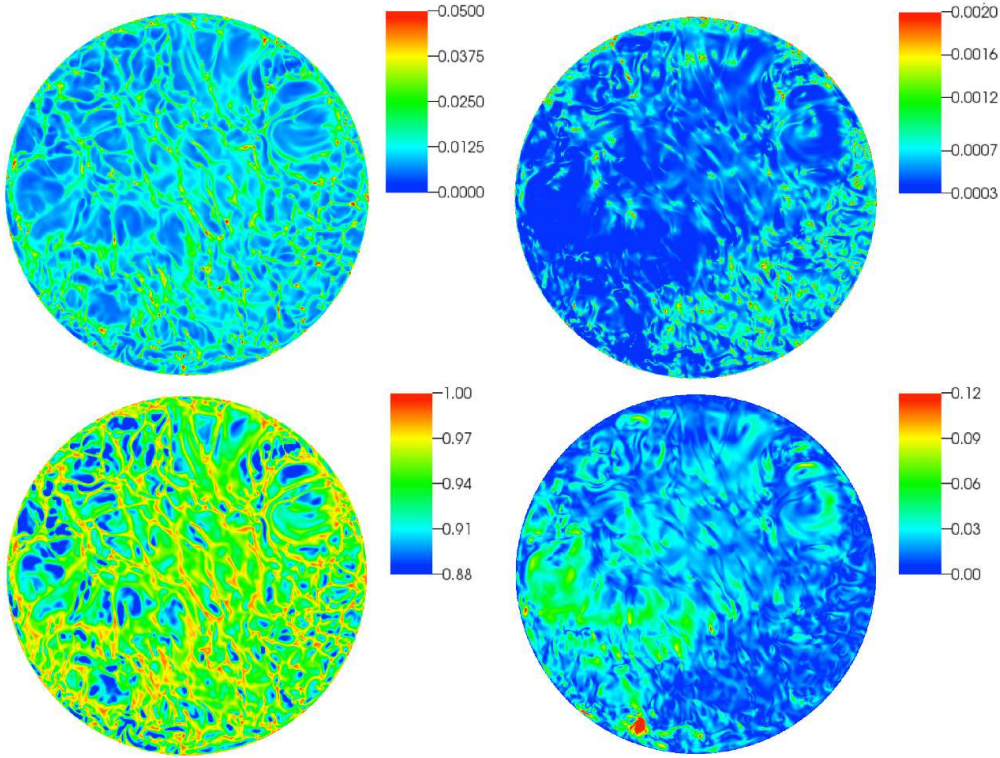}
\caption{Contour plots of snapshots of the local boundary layer scales and original turbulent fields. Top row: 
Snapshots of the outer local thermal boundary layer scale $\lambda^{o}_T$ (left) and the 
local inner velocity boundary layer thickness $\lambda^{i}_v$ (right). 
Both plots are taken at the bottom plate. Bottom row: Snapshots of temperature $T$ (left)   and the velocity magnitude (right). Both contour plots are
taken  at $z=0.0005$ for $Ra=10^{10}$, $\Gamma = 1.0$ and $Pr = 0.7$. All thickness scales are measured in units of $H$.}
\label{blpix}
\end{figure}
The strong variation of the local boundary layer scales becomes clearly visible in Fig. \ref{blpix} where we display contours of instantaneous plots of
$\lambda_T^{o}(x,y)$ and $\lambda_v^{i}(x,y)$. For comparison, we add the temperature and velocity fields to the figure from which the boundary layer scales have been derived. 
One sees that the local boundary layer scale of the temperature field (upper left) is well-correlated with the original temperature (lower left). Local maxima of $T$ which indicate
local detachments of thermal plumes are in line with enhanced local thicknesses.  Likewise the right column of  Fig. \ref{blpix} shows the results of 
using equation (\ref{velBL}) for the local inner velocity boundary layer scale (upper right) for the bottom plate. One sees that this boundary layer scale is anticorrelated 
with the magnitude of the horizontal velocity (lower right). The explanation for this behavior can be given based on the boundary conditions. Large horizontal velocities 
in the vicinity of the no-slip bottom boundary plane generate steep transversal velocity derivatives which cause a small inverse slope scale.

\begin{figure}
\centering
\includegraphics[width=\textwidth]{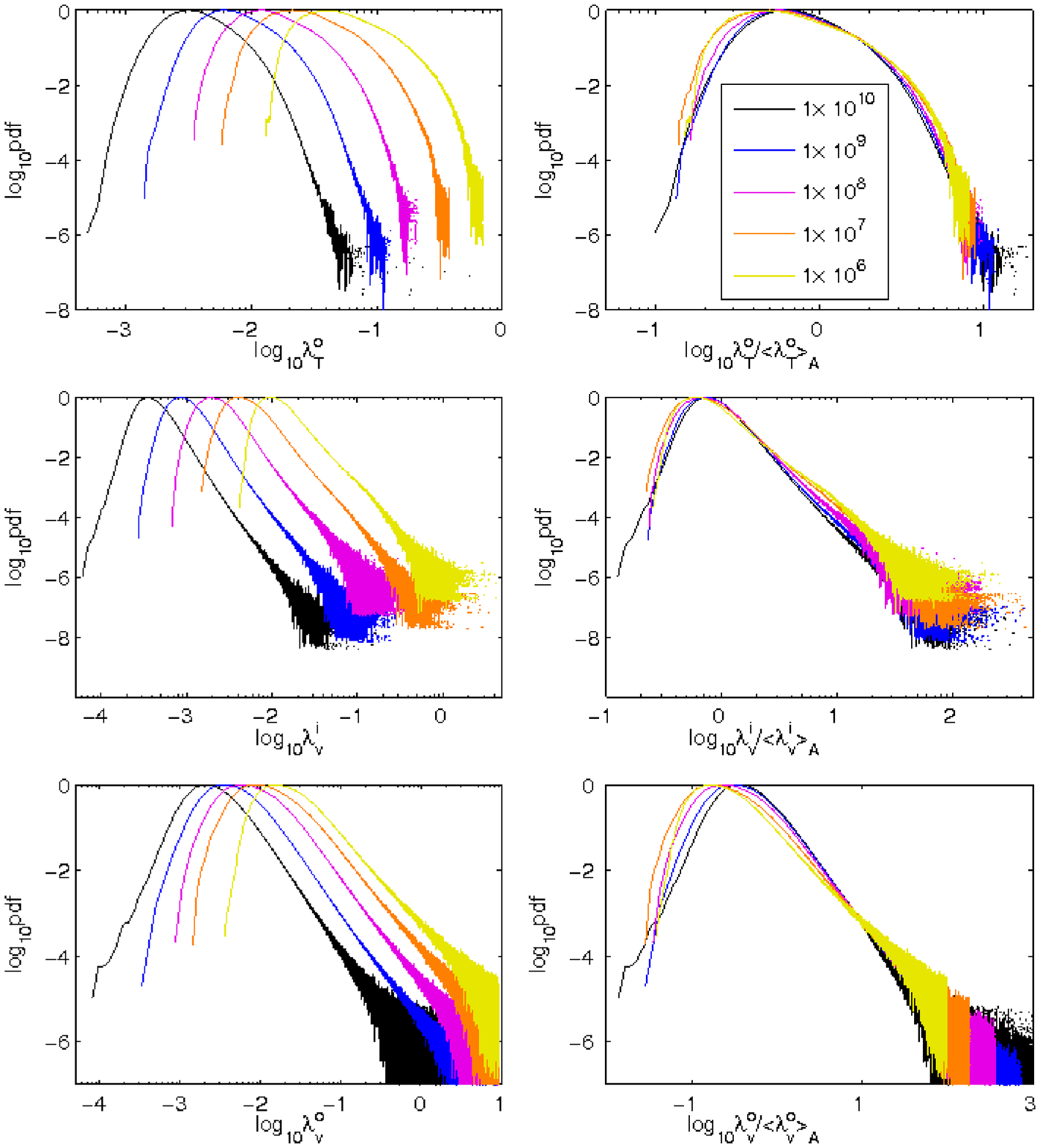}
\caption{Probability density functions (PDFs) of the outer local thermal boundary layer scale (top), the local inner velocity boundary layer thickness (middle) and the 
local outer velocity boundary layer thickness (bottom) for $\Gamma = 1.0$ and $Pr = 0.7$. The range of Rayleigh numbers is given in the legend. Data are obtained over 
a sequence of snapshots and the whole bottom and top plates. All thickness scales are measured in units of $H$.}
\label{blall}
\end{figure}
Next, we plot the unscaled results of the probability density functions (PDFs) for representative  Rayleigh numbers in the left column of Fig. \ref{blall}. 
These data are taken over the entire bottom and top plates and over a sequence of snapshots. Note the steady march towards smaller values of the BL thickness as the Rayleigh number increases. We will investigate the scaling laws for the mean values of
these distributions further below. Note also the fairly wide distribution in boundary layer thicknesses about the mean, which is consistent with what was found by \cite{Lui1998}. In the right column Fig. \ref{blall} we replot the PDFs, now scaled with their respective mean values as given in (\ref{tempBL1}), (\ref{velBL1}), and (\ref{velBL2a}). We denote the mean of the PDF obtained for the whole cross section planes of the cylindrical cell, $A$, by $\langle\cdot\rangle_{A}$.
Note the overall universal behavior and good collapse for both the local boundary layer scales of velocity and temperature over most of the range.
The PDFs always deviate from a lognormal distribution which would be perfectly symmetric with respect to $\lambda/\langle\lambda\rangle=1$ in a double-logarithmic plot. 
However $\lambda_v^o$ does not collapse as well as $\lambda_T^o$ or $\lambda_v^i$. 

The lower left plot in Fig. \ref{blall} deserves more comment, as one can clearly observe unphysical local outer velocity boundary layer thicknesses that are larger than $H$. This is a consequence of very shallow vertical gradients in ${\bf{u}}^{(2)}$ that can occur locally which lead to very large values of $\langle\lambda_v^o\rangle$. The multiplication by $u_{rms}$ in our definition (\ref{velBL1}) helps mitigate these shallow gradients, but still does not eliminate them. We have chosen to present the data as is, since locally these data still do correctly quantify the local boundary layer thickness right at the plate, and do not wish to impose additional prefactors or cutoff values.
\begin{figure}
\centering
\includegraphics[width=\textwidth]{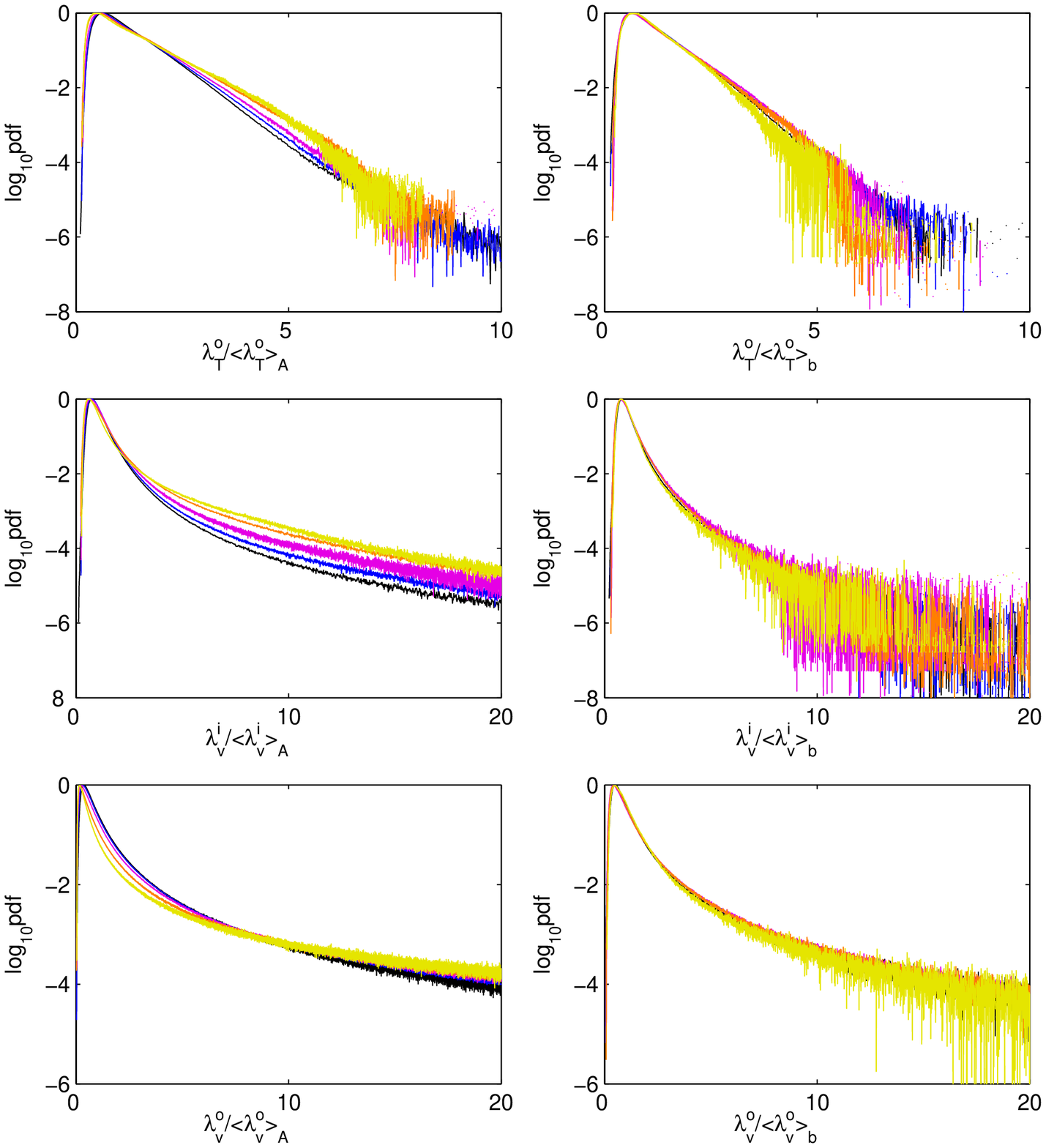}
\caption{PDFs of the local thermal boundary layer thickness (top row), the local inner velocity boundary layer thickness (middle) and the local outer 
velocity boundary layer thickness (bottom) for $\Gamma = 1.0$ and $Pr = 0.7$. The range of Rayleigh number is the same as in Figure \ref{blall}.  In this case, each PDF is 
scaled by its mean value. The left column shows the same data as in the right column of  Figure \ref{blall} (but on a semilogarithmic plot now) and the right column shows the PDF
results taken over a ``bulk'' region defined as radius $r < 0.3$, but still averaged over the bottom and top plates as well as the  time.}
\label{compbulk}
\end{figure}

We further investigate the local BL thickness PDFs by restricting our analysis to the ``bulk'' of the plates, where the bulk is defined as all of those BL thickness quantities computed for radii $r< 0.3$. This removes the sidewall effects. It is well-known (\cite{Lui1998}, \cite{Stevens2010}, \cite{Wagner2012}) that the boundary layer thicknesses differ between the center of the plate and the sidewalls because of the backrolls and jets at the sidewalls. Our results here support this. The first moment of the PDF obtained from these data is denoted as $\langle\cdot\rangle_{b}$. We show these PDFs as a function 
of the Rayleigh number scaled with the corresponding mean value in the right column of Fig. \ref{compbulk}. 
We normalize our bulk PDFs with the corresponding reduced plate area. Note that the PDFs taken over the entire plate (as shown in the left column) span a wider range of boundary layer thicknesses than their corresponding bulk PDFs. Also for the PDFs taken over the  whole plate there is a consistent trend towards smaller scales in the right tails as Rayleigh number increases. In contrast, for the PDFs taken over the bulk, the overall collapse is excellent.

\begin{figure}
\centering
\includegraphics[width=\textwidth]{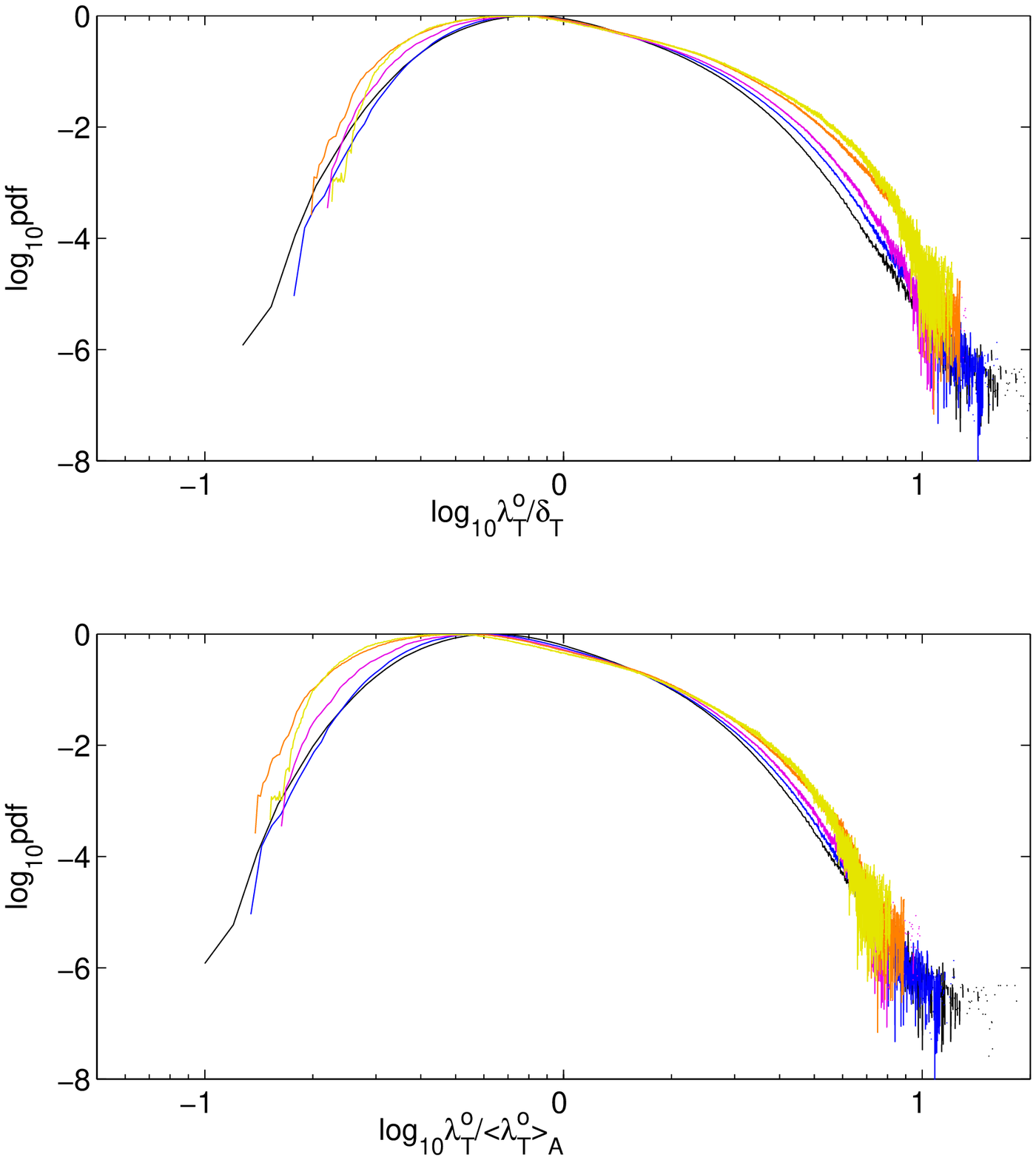}
\caption{PDFs of the local thermal boundary layer thickness where the mean value is obtained from (top) the Nusselt number (as listed in Tab. 1) and given by Eq. (\ref{TBL}) or (bottom) the 
average of the PDF as given by Eq. (\ref{tempBL1}) for $\Gamma = 1.0$ and $Pr = 0.7$. The range of Rayleigh numbers is the same as in Figure \ref{blall}. The data collapses better when scaled by 
the average of the PDF. Both analyses have been carried out with respect to the whole cross sections at top and bottom.}
\label{blscomnusvsmean}
\end{figure}

\begin{figure}
\centering
\includegraphics[width=1.0\textwidth]{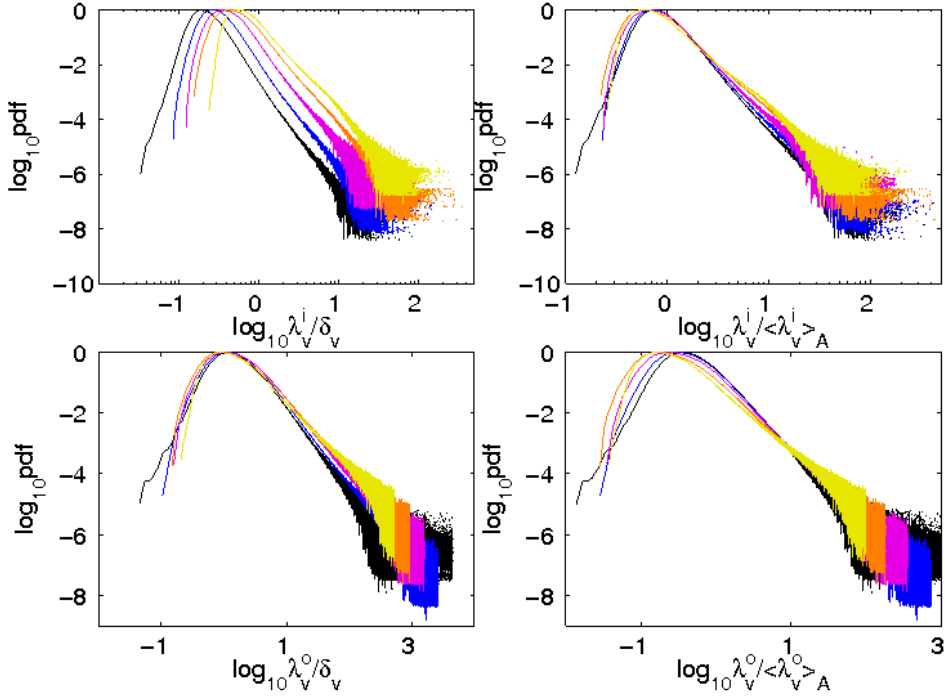}
\caption{PDFs of the inner (top) and outer local velocity boundary layer scale (bottom) where the mean value is obtained from (left column) the Reynolds 
number as listed in Tab. 1 and given by Eq. (\ref{vBL}) or (right column) the average of the PDF as given by Eq. (\ref{velBL1}) for $\Gamma = 1.0$ and $Pr = 0.7$. 
The range of Rayleigh numbers is the same as in Figure \ref{blall}.  Both analyses have been carried out with respect to the whole cross sections at top 
and bottom.}
\label{blvcomnusvsmean}
\end{figure}

To better understand the collapse of the data, we compare scaling of the $x$-axis of the PDFs with the mean value as computed from (\ref{tempBL1}) in the lower plot of 
Fig. \ref{blscomnusvsmean} with the mean value computed from the theoretical value for $\delta_T$ (see Eq. (\ref{TBL})) in the upper plot. Note the data collapse better for the 
scaling with the mean value of the PDF. However, the collapse is still good for the scaling with the theoretical value, except for a stronger trend towards smaller scales in the right tail for 
the largest Rayleigh numbers. The good collapse of the data suggests that our generalization to the outer local boundary layer scale of the temperature can be related consistently to the classical 
thermal boundary layer thickness Eq. (\ref{TBL}), i.e., $\langle \lambda_T^{o}\rangle \sim\delta_T$. In Fig. \ref{blvcomnusvsmean} we compare the collapse of the data for the local boundary layer 
scales of velocity. In the case of the inner scales, $\lambda_v^{i}(x,y)$, we see once more a good collapse in the center of the PDF when amplitudes are normalized by $\langle\lambda_v^{i}\rangle$.
As expected the collapse is not obtained when the inner scales are normalized by the Prandtl-Blasius type expression for $\delta_v$ found from Eq. (\ref{vBL}). As can be seen in the lower left panel of the figure, the opposite is 
the case for the outer slope-based scale. The data collapses now quite well for a rescaling by both, $\delta_v$ and $\langle\lambda_v^{o}\rangle$. The data also show that the collapse of
the distributions is significantly better for $\delta_v$. The coefficient in Eq. (\ref{vBL}) was taken to $a=1/4$ following \cite{Ahlers2009}. To conclude, the mean outer scale can be 
consistently related to the classical equation, i.e. $\langle \lambda_v^{o}\rangle \sim \delta_v$. Qualitatively, both scales display thus the same trend with Rayleigh number. 

\subsection{\label{Scalinglaws} Scaling of mean thickness scales with Rayleigh number}

\begin{figure}
\centering
\includegraphics[width=\textwidth]{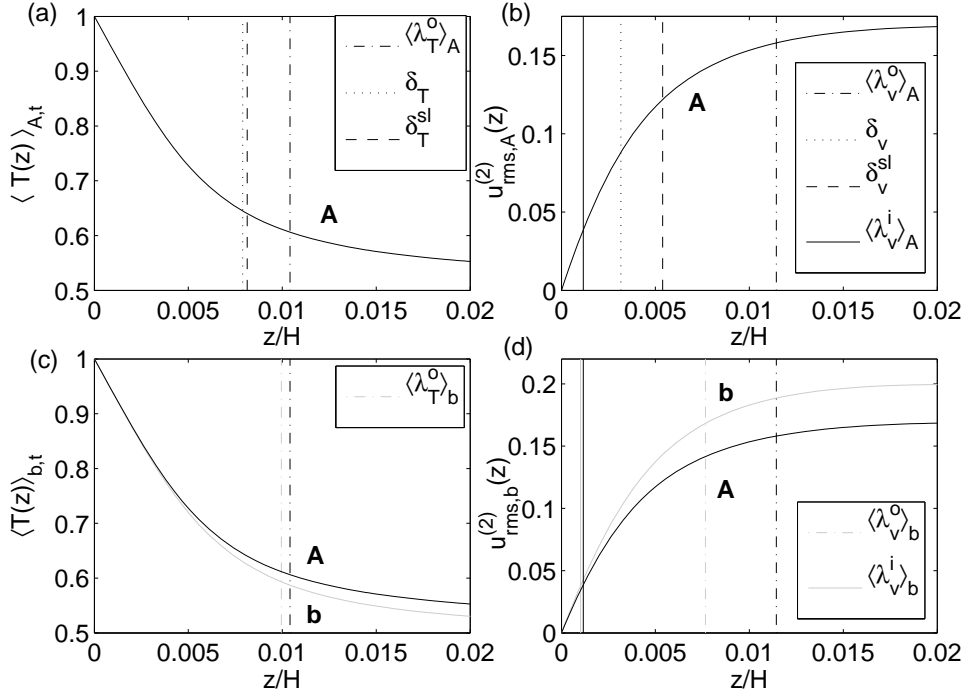}
\caption{A representative illustration of the mean boundary layer thickness scales for a Rayleigh number of $1\times 10^9$. (a) Plot of the 
time-averaged temperature profile averaged over the whole plate ($\langle T(z)\rangle_{A,t}$ (solid line)). The vertical lines indicate the location of the computed 
mean boundary layer thicknesses: the dashed-dotted line corresponds to $\langle \lambda_T^o\rangle_{A}$, the dotted line corresponds to  $\delta_T$, and the 
dashed  line corresponds to $\delta_T^{sl}$. (b) Plots of the time averaged profile $u_{rms, A}^{(2)}(z)$ as defined in Eq. (\ref{urms2}) (see solid line).
The vertical lines now correspond to: $\langle \lambda_v^o\rangle_{A}$ (dashed-dotted line), $\langle \lambda_v^i\rangle_{A}$  (solid vertical line), $\delta_v$ 
(dotted line) and $\delta_v^{sl}$ (dashed line). Panels (c) and (d) show some of the same data as the upper plots. All of the black lines are the same as in the corresponding 
panels (a) and (b), but the lighter gray lines indicate bulk-averaged quantities ($r<0.3$) instead.
\label{compprof2}}
\end{figure}

\begin{figure}
\centering
\includegraphics[width=0.72\textwidth]{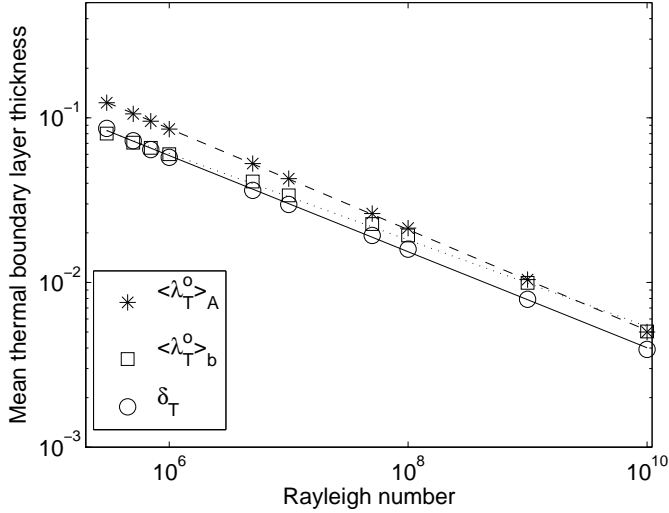}
\caption{Scaling of mean thermal boundary layer thickness with Rayleigh number for different analysis methods. The  
asterisks are for $\langle\lambda_T^o\rangle_{A}$ taken over the whole cell and with respect to time, and the fit is 
$\langle\lambda_T^o\rangle_{A} = (5.9 \pm 0.2)Ra^{-0.31\pm 0.01}$ (dashed line). The open squares are for  $\langle\lambda_T^o\rangle_{b}$ 
taken over the bulk volume with $r<0.3$, and the fit is $\langle\lambda_T^o\rangle_{b} = (2.3 \pm 0.3)Ra^{-0.26\pm 0.01}$ (dotted line). The 
open circles are the theoretical values for $\delta_T$ as defined in Eq. (\ref{TBL}) with the Nusselt numbers taken from Tab. 
1 and the fit is $\delta_T = (3.3\pm 0.3)Ra^{-0.29\pm 0.01}$ (solid line).}
\label{bltscal_comp}
\end{figure}
\begin{figure}
\centering
\includegraphics[width=0.9\textwidth]{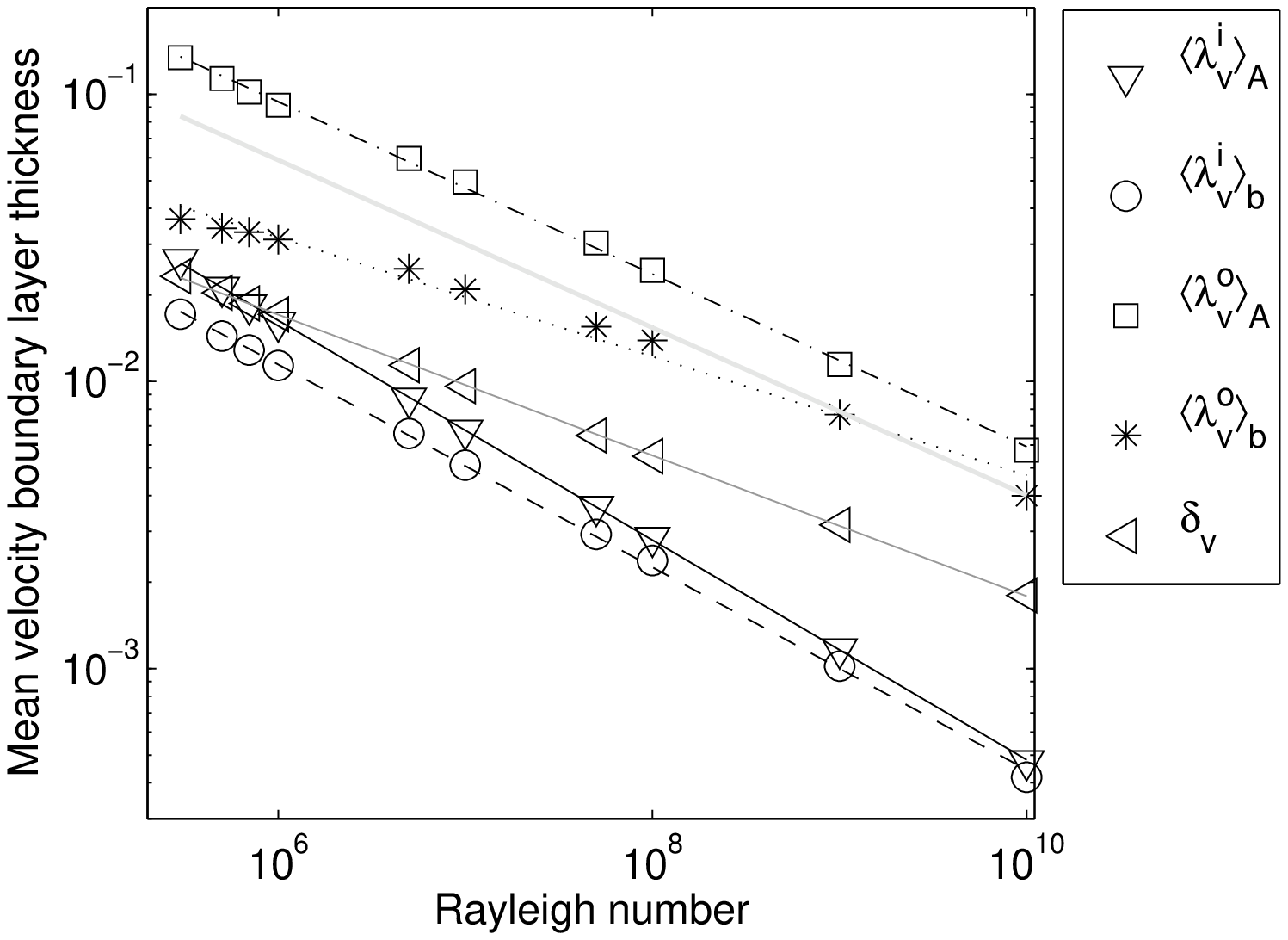}
\caption{Scaling of mean velocity boundary layer thickness with Rayleigh number for different analysis methods. 
The open downward directed triangles are for $\langle\lambda_v^i\rangle_{A}$ taken over the whole cell, and the fit is 
$\langle\lambda_v^i\rangle_{A}= (3.2 \pm 0.2) Ra^{-0.38\pm 0.01}$ (solid black line). The open circles are for  $\langle\lambda_v^i\rangle_{b}$ taken over 
the bulk, and the fit is $\langle\lambda_v^i\rangle_{b} = (1.5 \pm 0.2)Ra^{-0.35\pm 0.01}$ (dashed black line).  The open squares are for $\langle\lambda_v^o\rangle_{A}$ 
taken over the whole cell, and the fit is $\langle\lambda_v^o\rangle_{A} = (5.9 \pm 0.7)Ra^{-0.30\pm 0.01}$ (dash-dotted black line). The asterisks are for  
$\langle\lambda_v^o\rangle_{b}$ taken over the bulk, and the fit is $\langle\lambda_v^o\rangle_{b} = (0.55\pm 0.18)Ra^{-0.21\pm 0.01}$ (black dotted line). 
Note that the fit to the last six data points is $\langle\lambda_v^o\rangle_{b} = (1.0\pm 0.4)Ra^{-0.24\pm 0.01}$. 
The leftward pointing open triangles are the theoretical values for $\delta_v$ as defined in Eq. (\ref{vBL}) with the Reynolds numbers taken from Tab. 1 and $a=1/4$. 
The fit is $\delta_v = (0.50\pm 0.02)Ra^{-0.25\pm 0.01}$ (gray solid line). For comparison, we also added the fit for $\delta_T$ taken from Fig. \ref{bltscal_comp} as a solid 
bright gray line without symbols.}
\label{blvscal_comp}
\end{figure}

In the following we investigate the scaling laws for the mean boundary layer thickness scales with respect to Rayleigh  number. Relations  (\ref{tempBL1}), (\ref{velBL1}) and (\ref{velBL2a}) are 
applied. Representative plots of the location of the mean boundary layer thickness scales in relation to the mean temperature (left column) and mean velocity profiles (right column) are shown in 
Fig. \ref{compprof2}. For the velocity profiles, we used 
\begin{equation}
u_{rms, k}^{(2)} (z) = \sqrt{\langle u_x^2+u_y^2\rangle_{k,t}}
\label{urms2}
\end{equation}
where $k=\{A,b\}$. 
Vertical lines are drawn to highlight the location of the various boundary layer thicknesses. In the upper left plot, we show $\langle \lambda_T^o\rangle_{A}$ (dashed-dotted line) and $\delta_T$ (dotted line). We also plot $\delta_T^{sl}$ (dashed line), which is the boundary layer thickness found from the slope method for the plotted profile. Note that our calculated  $\langle \lambda_T^o\rangle_{A}$
 lies further from
$\delta_T$ than the value obtained from $\delta_T^{sl}$. 
Similar vertical lines are plotted in the top right panel for the velocity boundary layer thicknesses along with $\langle \lambda_v^i\rangle_{A}$ (solid vertical line).  One sees that the outer thickness scales are larger than the (appropriately named) inner thickness scales, and also that $\delta_v$ obtained from Eq. (\ref{vBL}) falls in between, as does $\delta_v^{sl}$. We see that again $\delta_v^{sl}$ agrees better with $\delta_v$ than our calculated $\langle \lambda_v^o\rangle_{A}$. The bottom panels show that the bulk-averaged profiles (solid gray curved lines) rise more steeply than the profiles obtained from averaging over the entire area (solid black curved lines). This is why we tend to obtain smaller bulk-averaged boundary layer thicknesses (vertical gray lines), which are shown on the plots along with the corresponding area-averaged quantities (vertical black lines). Whereas for the thermal boundary layer thicknesses, all values are quite close to one another, the bulk-averaged quantity $\langle \lambda_v^o\rangle_{b}$ agrees better with $\delta_v$ than the plate-averaged outer boundary layer thickness  $\langle \lambda_v^o\rangle_{A}$. We chose to plot the raw data to highlight the steeper rise and correspondingly smaller boundary layer thicknesses for the bulk-averaged quantities.

The scaling results are summarized in Fig. \ref{bltscal_comp} for the temperature and in
Fig. \ref{blvscal_comp} for the velocity. The corresponding classical values for the BL thicknesses are given by (\ref{TBL}) and  
(\ref{vBL}), respectively. Again, we took $a=1/4$ for $\delta_v$.

The agreement in Fig. \ref{bltscal_comp} between $\langle\lambda_T^o\rangle_A$ and $\langle\lambda_T^o\rangle_b$ 
is fairly good, especially for larger Rayleigh numbers. We conjecture that this 
convergence is caused by the increasing filamentation of the thermal plumes as the Rayleigh number grows. As a larger number of finer plumes are
distributed across the plane, the side wall regions become increasingly less important. Consequently, the results obtained for planes 
$A$ and $b$ converge. Conversely for the lower Rayleigh numbers, the large-scale circulation and corresponding sidewall backrolls and jets are comparatively stronger so one expects a larger difference between planes $A$ and $b$. The classical boundary layer thickness scaling is always a bit lower than our mean local boundary layer thicknesses, with a larger discrepancy at lower Rayleigh number and when averaging over $A$ instead of $b$. We find that the steep vertical gradients of the temperature field at the plate (the pdf of $dT/dz$ is highly skewed towards larger magnitudes)  cause a larger $Nu$ and hence smaller $\delta_T$. Whereas when performing the average in (\ref{tempBL1}), the inverse of those steep gradients make  very little contribution to $\langle\lambda_T^o\rangle$ and hence our mean local boundary layer thicknesses tend to be on average a bit larger. This effect is enhanced when averaging over $A$ instead of $b$.
\begin{table}
\begin{center}
\begin{tabular}{ l  c  c   c  c  c }
Group & Range of $Ra$ & $\alpha_{T}$ & $\beta_{T}$ & $\alpha_{v}$ & $\beta_v$ \\ \hline
Current work* & $10^5-10^{10}$ & 2.3$\pm$0.3  & -0.26$\pm$0.01 & 0.55$\pm$0.18 &  -0.21$\pm$0.01\\ 
\cite{Scheel2012}* & $10^5-10^8$ & 1.76$\pm$0.12 & -0.25$\pm$0.01 & 0.40$\pm$0.14 & -0.18$\pm$0.01 \\
\cite{Li2012} & $10^{9}-10^{12}$ & 0.42   & -0.24   & 0.90 & -0.24 \\  
\cite{Wagner2012}* & $10^{4}-10^{9}$ & (---)   & -0.285 $\pm$ 0.003  & (---) & -0.238 $\pm$ 0.009\\ 
\cite{duPuits2007b} & $10^9-10^{12}$ & (---) & -0.2540 & (---)&  (---) \\ 
\cite{Verzicco1999}* & $10^{5}-10^{8}$ & 3.1  & -0.29 & 0.95 & -0.23 \\
\cite{Belmonte1994} & $10^7-10^{11}$ & 2.5 & -0.29 & (---) & (---) \\   \hline
\end{tabular}
\end{center}
\caption{Comparison of scaling coefficients for the thermal and velocity
boundary layer thicknesses (see Eq. (\ref{compcoeff}))}. For the current work we selected the scaling for $\langle\lambda_T^o\rangle_b$ and likewise $\langle\lambda_v^o\rangle_b$ 
since these most closely resemble the methods used in the rest of the referenced works. We only selected cases for $Pr \simeq 0.7$, $\Gamma \simeq 1$ 
and the Rayleigh number range (given) which overlapped with our data range. The asterisks indicate numerical simulations; the others are experiments.
\label{tab:scaling} 
\end{table}

The analysis for the velocity in Fig. \ref{blvscal_comp} deserves a closer inspection. First, the outer scales over the whole 
plane $A$ or the bulk $b$ are larger than the classical thickness scales $\delta_v$ as was also seen in Fig. \ref{compprof2}. This indicates some freedom which is always left 
in the definitions due to different possible choices of the velocity. Interestingly, the scaling exponent of $\langle\lambda_v^{o}\rangle_A$  agrees better with $\delta_T$ 
than $\delta_v$ which suggests that performing an average over the entire plate for the velocity BL thickness is problematic.
Second, the mean scales averaged across the whole plane are significantly larger than the mean scales averaged across the bulk, especially for lower Rayleigh numbers. Third, the mean bulk scales $\langle\lambda_v^{o}\rangle_b$ decrease slower 
than the mean scales across the whole plane, $\langle\lambda_v^{o}\rangle_A$. 
We conclude that the recirculation processes close to the sidewalls which enhance the local boundary layer thickness become increasingly less dominant. 
 As a consequence, a convergence of both sets of scales is observed, similar to the outer local thermal boundary scales, $\langle\lambda_T^{o}\rangle_A$ and 
$\langle\lambda_T^{o}\rangle_b$. Our finding of a thicker local boundary scale close to the side walls is consistent with the result obtained by \cite{Lui1998} and \cite{Wagner2012} 
(see e.g. their Fig. 19). It is clear that these 
effects become more pronounced for smaller aspect ratio cells, and suggests that any BL analysis for $Ra < 10^{10}$ needs to carefully consider where to perform averages over the plates. 
We finally note that the inner scales for the velocity field yield 
consistently significantly smaller values than $\delta_v$.

We compare our scaling results with previous published results in Table \ref{tab:scaling}, where the coefficients are given by
\begin{equation}
\delta_T = \alpha_TRa^{\beta_T} \qquad \delta_v = \alpha_vRa^{\beta_v}
\label{compcoeff}
\end{equation}
 Since the coefficients tend to be sensitive to the particular parameters ($Ra, Pr, \Gamma$) we only selected experiments and numerical simulations which most closely resembled the current analysis. We also selected to compare only our coefficients for the scaling of the outer thicknesses. This most closely agrees with the slope method which was used in the other works. Since the spatial averaging in most of the other works either was local or removed the sidewall area (except for \cite{Verzicco1999}), we also chose to use our bulk averaged values.  We find that the overall agreement in the table for the exponents $\beta$ is fairly good. 
If we use only the last six data points ($Ra \geq 1\times 10^7$) for the fit to $\langle\lambda_v^o\rangle_{b}$ in Fig. \ref{blvscal_comp} we obtain $\langle\lambda_v^o\rangle_{b} = (1.0\pm 0.4)Ra^{-0.24\pm 0.01}$, which gives even better agreement with the other works in Table \ref{tab:scaling}. We find that the agreement with the coefficients $\alpha$ is not as satisfactory. The disagreement for both $\alpha$ and $\beta$ may in part be explained by the subtle differences in the extraction of boundary layer thicknesses for all of the compared cases and suggests that the coefficients (especially $\alpha$) are much more sensitive to the details of the measurements. Differences could also be attributed to the slightly different Rayleigh number ranges used.

\begin{figure}
\centering
\includegraphics[width=0.72\textwidth]{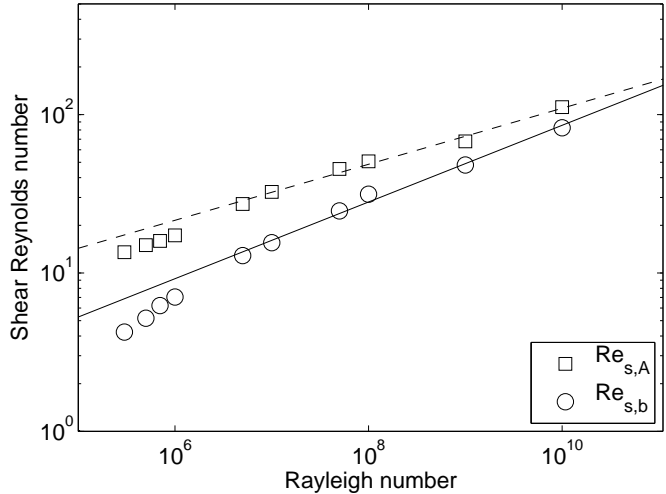}
\caption{Scaling of shear Reynolds number with Rayleigh number. The corresponding power law fits to the first six data points  are $Re_{s,A} = 
(1.9 \pm 0.6)Ra^{0.18\pm 0.01}$ (dashed line) and $Re_{s,b} = (0.3\pm 0.1)Ra^{0.24\pm 0.01}$ (solid line). The shear Reynolds numbers are 
determined in correspondence with Eqn. (\ref{ResA}).}
\label{Rescal_comp}
\end{figure}

What are the consequences of our mean thickness scales for the scaling of the shear Reynolds number $Re_s$ with respect
to the Rayleigh number $Ra$? We define the shear Reynolds numbers for the whole cell and the bulk as 
\begin{equation}
\label{ResA}
Re_{s,k}=\sqrt{\frac{Ra}{Pr}} \,\left[u_{rms, k}^{(2)} (z)\Big|_{z=\langle \lambda_v^{o}\rangle_{k}} \langle \lambda_v^{o}\rangle_{k}\right]\,,
\end{equation}
where $u_{rms, k}^{(2)}(z)$ its given by Eq. (\ref{urms2}) and $k=\{A,b\}$. Figure \ref{Rescal_comp} displays the shear Reynolds numbers as a function of the Rayleigh 
number and the corresponding fits which have been obtained for the six largest Rayleigh numbers in each of the two data sets. 
We note that the scaling exponent of the mean thickness obtained for the bulk $b$ is larger than the one for the whole area $A$ which underlines the 
substantial influence of the side wall regions, even for an aspect ratio $\Gamma=1$. Averaging over the entire plate continues to be problematic here, just as in Fig. \ref{blvscal_comp}.
Our bulk-averaged data agrees with \cite{Li2012} ($Re_s \propto Ra^{0.267\pm0.0386}$)
who determined the average by profile measurements at centerline of the cell. However, our scaling exponents are smaller and prefactors larger than those obtained by \cite{Wagner2012} ($Re_s = 0.072Ra^{0.2675}$), who also used a local (slope) method and a bulk average, but used the most probable boundary layer thickness instead of the mean value as  in Eq. (\ref{velBL1}). The most probable boundary layer thickness is smaller in all cases (see Figure \ref{blall}) than the first moment, which may explain the difference in coefficients. Our Rayleigh number range is also higher by a factor of 10, which could also account for the differences.

The line of thought we 
followed here is to obtain the mean thickness as a mean of local slope scales and to combine this with a mean horizontal velocity at a distance from the wall 
which is exactly given by this mean. This gives us a shear Reynolds number  in such a  way that no free parameter is left in our analysis. 

\subsection{Local boundary layer scales for different aspect ratios}
\begin{figure}
\centering
\includegraphics[width=0.9\textwidth]{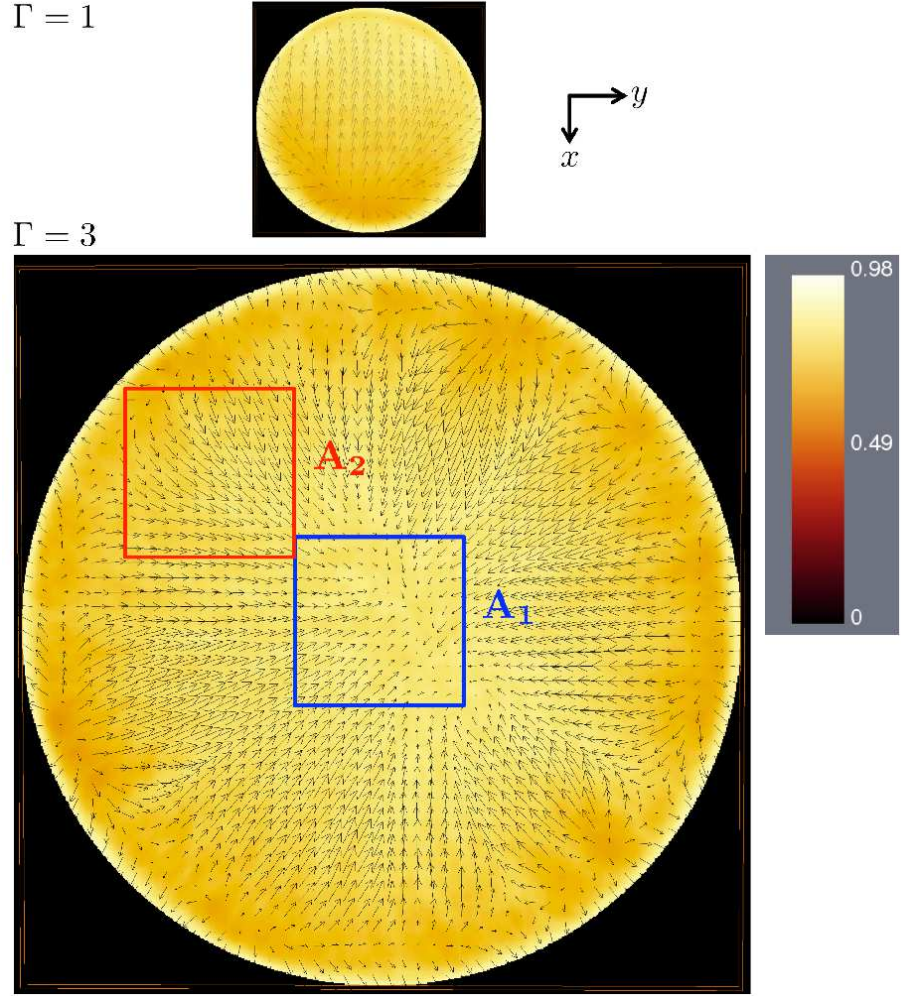}
\caption{Time averaged velocity and temperature field for the simulation runs at $Ra=10^8$. Top figure is for $\Gamma=1$, the bottom figure is for $\Gamma=3$.
We display a contour plot of the time-averaged temperature field at $z=\delta_T/2$ in combination with the time-averaged velocity field at $z=2\delta_T$.
Vectors are a projection into the $x$--$y$ plane. The sub areas $A_1$ and $A_2$ are used for a local analysis of the local boundary layer scales. Area $A_1$
is chosen at the interface between two large-scale circulation rolls while $A_2$ is inside one of the two large-scale circulation rolls. Both cases are a view from
the top onto the bottom region of the convection cell. All thickness scales are measured in units of $H$.}
\label{lsc}
\end{figure}

The analysis has been discussed so far for a cell with unity aspect ratio. In this case the large-scale circulation consists of a single roll. 
The question which we want to investigate in the following is if the boundary layer analysis is sensitive to the multi-roll circulation patterns that appear in larger aspect ratio systems 
(see \cite{Bailon2010} for a detailed discussion of the evolving large-scale patterns as a function of 
the aspect ratio). We conducted a DNS at an aspect ratio $\Gamma=3$ for a Rayleigh number $Ra=10^8$. In this setting two large circulation 
rolls co-exist. Figure \ref{lsc} compares the time-averaged temperature and horizontal velocity fields for $\Gamma=1$ and 3. We have averaged over 39 free-fall times for $\Gamma = 3$, short enough that the pattern has not evolved/drifted significantly enough to smear out the large-scale flow. For $\Gamma = 1$, we took this average over 104 free-fall times. Both time intervals are 
too short to capture the very slow dynamics of the large-scale circulation (\cite{Shi2012}). This would require significantly longer run times which we cannot perform for these fine grids. Therefore the mean flow 
which we display in both panels is quasi static to a very good approximation. Clearly observable
is the fingerprint of the single circulation roll in the top panel of the figure. This circulation which is downwelling at the bottom of the panel and upwelling at the 
top of the panel is in line with slight mean temperature differences across the plane which are taken at $z=\delta_T/2$. The colder spot of averaged 
temperature (brown) is connected with the downwelling toward the bottom plate, the warmer spot (yellow) with the upwelling from the bottom plate.

The mean flow pattern is more complex in the case of the larger aspect ratio. Near the center we observe a pronounced convergence zone where the 
circulation flow is upwelling. Colder spots of downwelling mean circulation are distributed close to the side walls. In a streamline plot (not shown) of the time-averaged velocity field one observes two circulation rolls. The boundary analysis for this cell is conducted either across the whole plane as before or it is localized in two windows which
are added to the lower panel of Fig. \ref{lsc}. Window $A_1$ ($-0.35 < x,y < 0.35$) is put right into the convergence zone with dominantly upward motion 
while window $A_2$ ($-1.05 < x < -0.35$ and $-1.00 < y < - 0.30$) corresponds with the situation as being present in the case of $\Gamma=1$. 
\begin{figure}
\centering
\includegraphics[width=\textwidth]{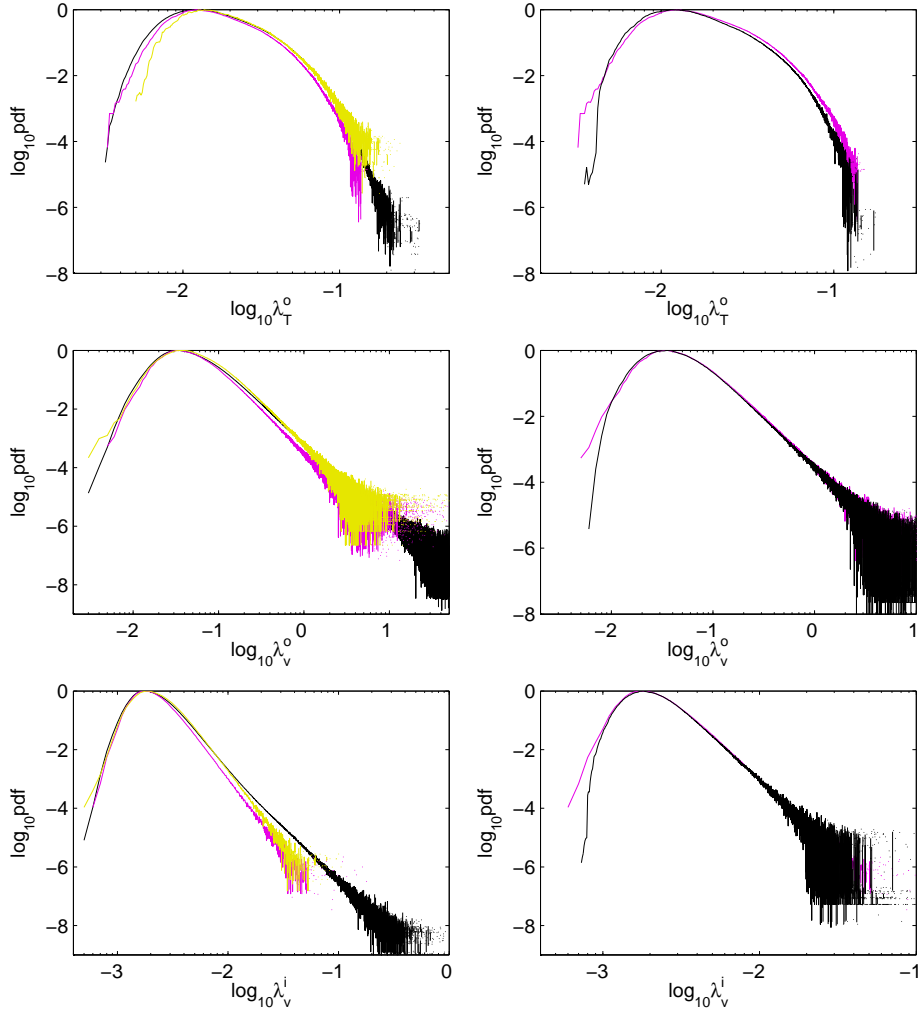}
\caption{A comparison of the PDFs of the local boundary layer methods for two different aspect ratios, (both at a Rayleigh number of $1\times 10^8$). 
The left panel compares the PDFs taken over the sub areas to PDFs of the entire cell all for $\Gamma = 3$ where: black is for the entire cross-sectional area, yellow (light gray) is $A_1$ and purple (dark gray) is
 $A_2$, where the sub areas are defined in Figure \ref{lsc}. The right panel compares the PDFs taken over the sub area $A_2$ to PDFs of the bulk for $\Gamma = 1$ 
 where: purple (dark gray) is $\Gamma = 3, A_2$ and black is $\Gamma = 1$, bulk ($r< 0.3$). All thickness scales are measured in units of $H$.}
\label{diffgamma}
\end{figure}

In Fig. \ref{diffgamma} we compare different PDFs for two different aspect ratios. Since the Rayleigh number is the same for the two different aspect ratios we chose here to not scale the boundary layer thicknesses by their corresponding mean values. The mean boundary layer thicknesses are given in Table \ref{tab:meancomp} for comparison and we see that the mean values for all  boundary layer thicknesses are smaller in the $\Gamma = 1$ cell than for the whole $\Gamma = 3$ cell. This is particularly true for the outer velocity boundary layer thickness which can be explained in part by the larger $u_{rms}$ prefactor for the $\Gamma = 3$ cell (see Table \ref{Tab1}). The other discrepancies can be explained by the different flow geometries, and when the average is taken over subvolume $A_2$, the $\Gamma = 3$ data agrees best with the $\Gamma = 1$ case as would be expected.

 In the left column of Fig. \ref{diffgamma}, data for $\Gamma=3$ taken over different areas are compared. Note that the local outer boundary layer thicknesses are larger than $H$ in some cases. This is for the same reason as in Figure \ref{blall} and we again present the data unscaled here. The distribution of $\lambda_T^{o}$ taken over $A_2$ shows the sparsest tail. The right tail of the PDF taken over the convergence zone $A_1$ is slightly fatter
as larger thermal plumes will prominently rise here and enhance the local thermal boundary layer scale. The PDF obtained for the whole cell falls consistently in between
which indicates that side wall effects become increasingly subdominant. The local analysis is almost insensitive with respect to the outer velocity scale, 
$\lambda_v^{o}$. The right tail of the data obtained over $A_2$ is slightly sparser. For the inner scale, $\lambda_v^{i}$, data obtained over $A_1$ and $A_2$ almost
coincide. Here, we observe a fatter tail for the analysis across the whole area. This circumstance might display the impact of recirculation zones at the side walls for which 
local wall shear stresses can change sign and therefore contribute to larger local scale events. 

In the right column of the same figure, the results obtained for $\Gamma=1$ in the bulk with $\Gamma=3$ for $A_2$ are compared. At a first glance, 
the PDFs of both data records coincide quite well in all three cases for the majority of amplitudes. We observe in all cases a more pronounced left and right tail for
the larger aspect ratio (except for $\lambda_v^i$ where the right tails almost coincide). Finer local scales imply steeper local gradients and vice versa. These gradients 
can be established since the dynamics in the convective turbulence is less constrained by cell 
geometry such that, e.g.,  the fluid can be swept over longer distances. Turning back to Fig. \ref{lsc}, it can be seen that the area $A_2$ with a nearly uniform mean flow is
almost as large as the whole area in the top panel of the same figure. Our conjecture might also be supported by the fact that the velocity fluctuations are larger in the case
of $\Gamma=3$ compared to $\Gamma=1$ at the same Rayleigh number of $Ra=10^8$, as documented in Tab. \ref{Tab1}. In summary, our  
analysis confirms that the results of the local boundary layer scale analysis is somewhat dependent on the geometry, mainly for the tail events which are 
associated with very fine or very coarse local scales.

\begin{table}
\begin{center}
\begin{tabular}{ l   c  c  c }
Area Range & $\langle\lambda_T^o\rangle$ & $\langle\lambda_v^o\rangle$ & $\langle\lambda_v^i\rangle$ \\ \hline
$\Gamma = 1$, $r<0.3$ & 0.0193$\pm$0.0001 & 0.0139$\pm$0.0002 & 0.00238$\pm$0.00001 \\
$\Gamma = 3$, whole plate & 0.0214$\pm$0.0001 & 0.0197$\pm$0.0002 & 0.00251$\pm$0.00001 \\
$\Gamma = 3$, $A_1$ & 0.0204$\pm$0.0001 & 0.0193$\pm$0.0003 & 0.00257$\pm$0.00001 \\
$\Gamma = 3$, $A_2$ & 0.0207$\pm$0.0001 & 0.0160$\pm$0.0002 & 0.00236$\pm$0.00001 \\\hline
\end{tabular}
\end{center}
\caption{Mean values of the various boundary layer thicknesses for different aspect ratios and different regions. The corresponding PDFs are plotted in Fig. \ref{diffgamma}.}
\label{tab:meancomp} 
\end{table}

\subsection{Dissipation layer thickness analysis}
\begin{figure}
\centering
\includegraphics[width=0.49\textwidth]{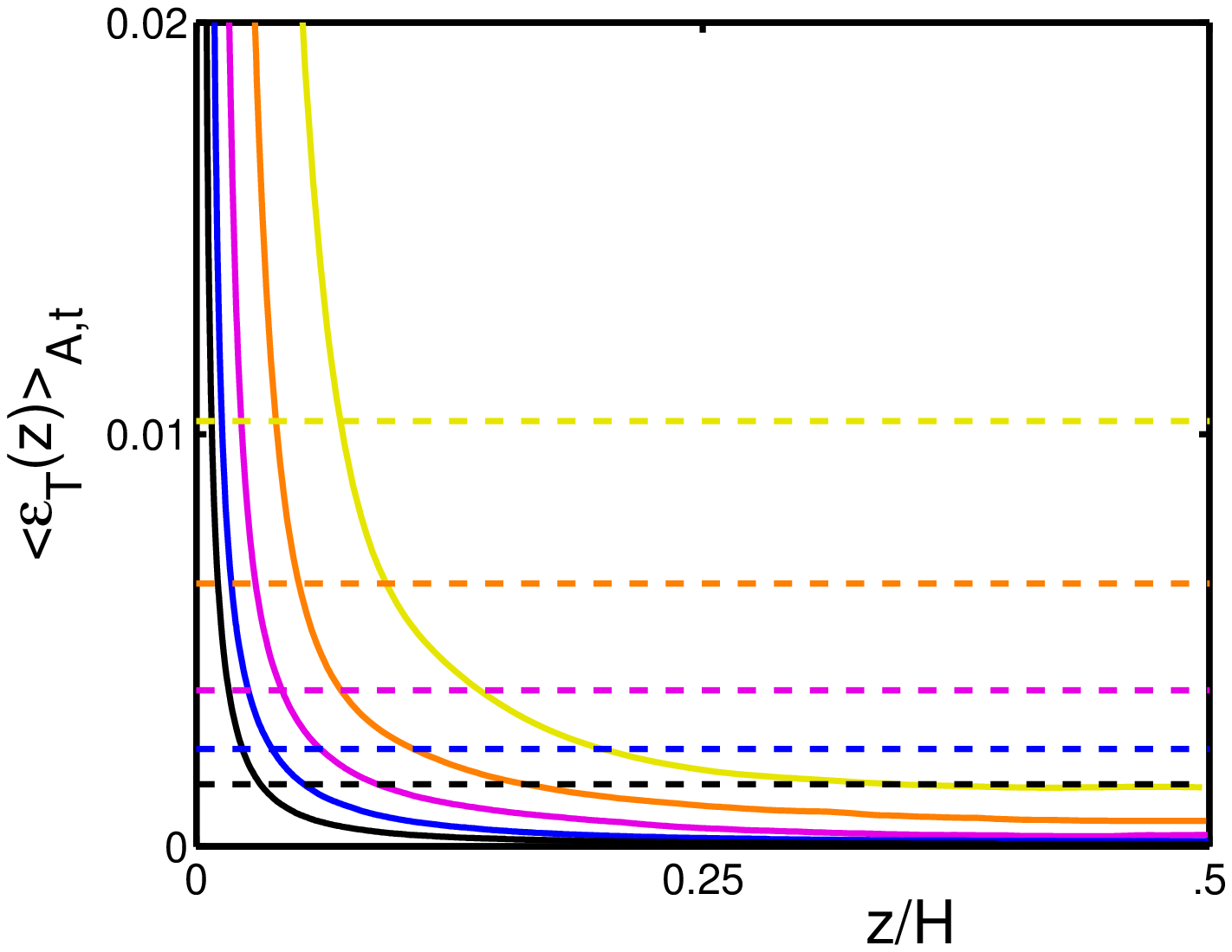}
\includegraphics[width=0.49\textwidth]{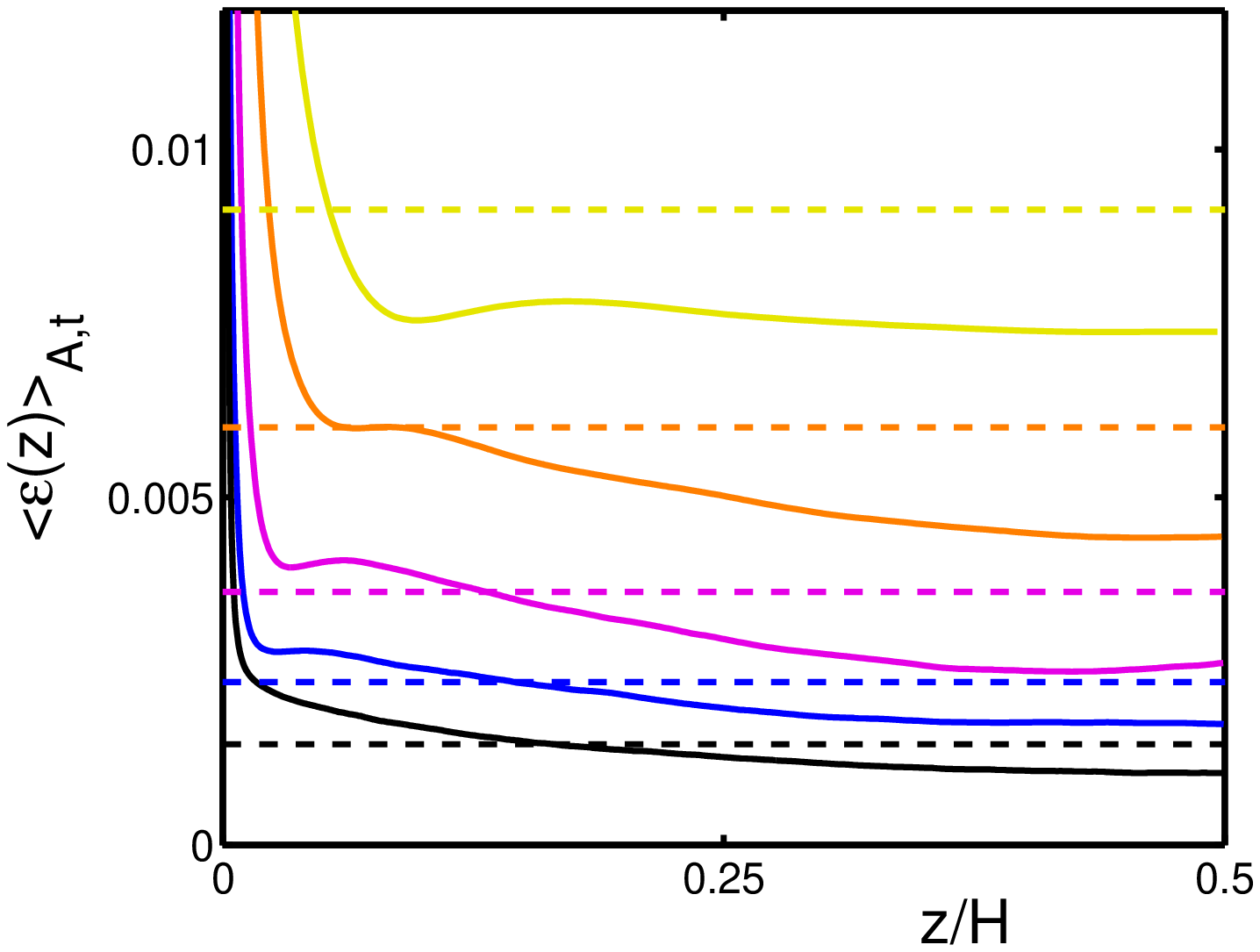} 
\caption{Horizontally averaged dissipation rates (solid lines) for various Rayleigh numbers. The left panel is for  $\langle\epsilon_T(z)\rangle_{A,t}$ and 
the right panel is for $\langle\epsilon(z)\rangle_{A,t}$. The profiles were also averaged about $z/H=0.5$ so that the top and bottom half both contribute to 
the profile. The dashed lines give the volume averaged dissipation rate $\langle\epsilon_T\rangle_{V,t}$ (left) and $\langle\epsilon\rangle_{V,t}$ (right). 
The intersection of these two lines  gives the  dissipation layer thickness  $d_{T}$  and $d_{v}$, respectively.  The range of Rayleigh number is given  in 
Fig. \ref{blall} and the parameters are the same. }
\label{bltfromdiss}
\end{figure}

Following \cite{Petschel2013}, we compute the velocity and thermal dissipation layer thicknesses from (\ref{disslayer}).
As already mentioned in the introduction, we will denote the kinetic energy dissipation layer thickness as $d_v$ and the thermal dissipation layer thickness as $d_T$.
While \citeauthor{Petschel2013} kept the Rayleigh number at a moderate value of $Ra=5\times 10^6$ and varied the Prandtl number over
a very wide range $(0.01\le Pr\le 300)$, we will use our data to study the dissipation scales as a function of Rayleigh number. We show representative examples 
of dissipation layer thickness  calculations in Fig. \ref{bltfromdiss}. In the left panel, the plane-time averaged thermal dissipation
profiles (solid lines) and the corresponding volume mean values (dashed lines) are displayed. The same is repeated for the kinetic energy dissipation
rate in the right panel. We observe a systematic trend in the case of $d_T$. The mean thermal energy dissipation rate decreases and the intersection point
with the vertical profile is shifted in proportion towards the wall indicating an increasingly finer thermal dissipation layer thickness $d_T$.

\begin{figure}
\centering
\includegraphics[width=0.8\textwidth]{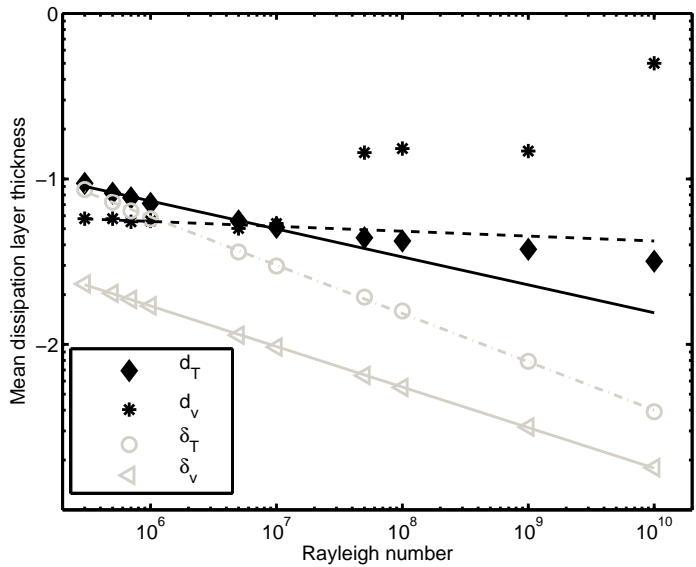}
\caption{Scaling of thermal and velocity dissipation layers with Rayleigh number. The black solid diamonds are for $d_T$, and the fit to the first six data points (solid black line) 
is $d_T = (0.8\pm 0.2)Ra^{-0.17\pm 0.01}$. The black stars are for $d_v$, and the fit to the first six data points (dashed line) is $d_v = (0.08\pm 0.02)Ra^{-0.03\pm 0.01}$. Note the interesting transition for 
the dissipation layer thicknesses for $1\times 10^7 < Ra < 1\times 10^8$. The classical boundary layer thicknesses and their corresponding fits are given by the grey 
symbols and lines (same data as in Figures \ref{bltscal_comp} and \ref{blvscal_comp}).}
\label{blfromdisscal}
\end{figure}

The situation is different in the case of the kinetic energy dissipation layer thickness.  Again, the mean energy dissipation decreases
with increasing Rayleigh number. However the mean vertical profile changes: in particular the dissipation starts  to decrease monotonically 
towards the mid plane for the higher Rayleigh numbers. This causes a jump in the intersection point which becomes visible in 
Fig. \ref{blfromdisscal}. This figure shows the dissipation layer thickness data for all of the Rayleigh numbers we used in our study. We 
observe a change in the scaling at $Ra\sim 10^7$ for both the thermal and the kinetic dissipation layer thicknesses. While the thermal dissipation layer thickness changes from a power law exponent
of -0.17 for the first six data points  to -0.07 for the last five data points, the kinetic energy dissipation layer thickness remains almost 
constant (we fit an exponent of -0.03 to the first six data points) initially and increases afterwards non-monotonically. As a guide to the 
eye we added the classical thickness  data of $\delta_T$ and $\delta_v$ to Fig. \ref{blfromdisscal} as gray lines and symbols.

At this point we can only speculate about the reasons for a change in the trend with $Ra$, in particular for the kinetic energy 
dissipation layer thickness. We conjecture that this transition is related to a transition observed by \cite{Schumacher2014}. They found that the scaling of the moments 
of the kinetic energy dissipation rate (defined in Eq. (\ref{kinetic1})) with Reynolds number follow universal power laws for sufficiently high Reynolds numbers. Hence, 
beyond the transitional Reynolds number, small-scale turbulence in the bulk obeys the properties of a fully developed turbulent state. For Rayleigh-B\'enard convection 
this transition occurs at a Reynolds number corresponding to $Ra\simeq 5\times 10^6$.
Table \ref{Tab1} indicates that the rms values of the velocity field ${\bf u}$ remain at about the same level up to a Rayleigh 
number  of $10^7$ and start to decrease slowly but monotonically with increasing Rayleigh number. The mean energy  dissipation is found 
to follow a power law of $\langle\tilde{\epsilon}\rangle_{V,t}=(0.13\pm 0.02)\times Ra^{-0.19\pm 0.01}$ which is also consistent with the scaling reported in \cite{Scheel2013},
obtained there over a smaller range of Rayleigh numbers than in the present work. The decrease of the velocity fluctuations can 
be related to a reduced  coherence of the detaching thermal plumes. Consequently the mean vertical dissipation profiles will vary because
strong shear layers due to rising or falling plumes becomes less pronounced. 
Our simulation data suggest that both dissipation layer thickness
scales pass through a transition in their Rayleigh number dependence.

\subsection{Friction coefficient}\label{cepschapter}
\begin{figure}
\centering
\includegraphics[width=0.72\textwidth]{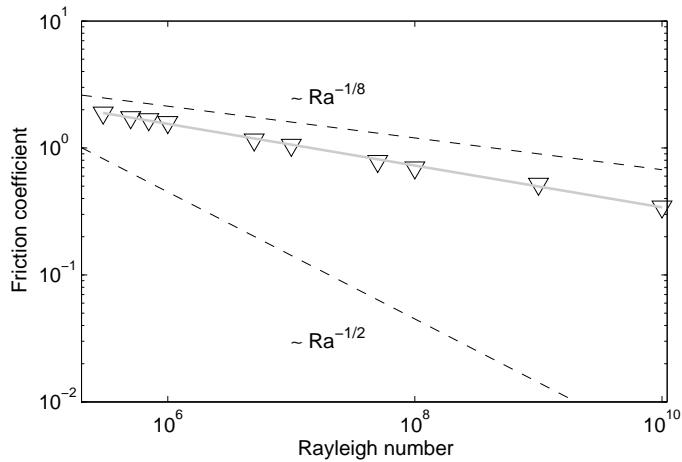} 
\caption{Friction coefficient $c_{\epsilon}$ as a function of the Rayleigh number. As a guide to the eye we add the scaling which for a purely laminar flow is $c_{\epsilon}\sim Ra^{-1/2}$,
and for a fully developed turbulent wall-bounded flow is  $c_{\epsilon}\sim Ra^{-1/8}$ (dashed black lines). These scalings would follow when the Reynolds number is 
related to the Rayleigh number by $Re\sim\sqrt{Ra}$. The fit to the simulation data gives $c_{\epsilon}=(15.0\pm 1.4) \times Ra^{-0.16\pm 0.01}$
 and is indicated by a solid gray line.}
\label{c_eps}
\end{figure}

A further interesting question which is related to the discussion in the last subsection is how the friction coefficient behaves in the convection 
system. We define here the dimensionless friction coefficient as
\begin{equation}
\label{ceps}
c_{\epsilon}=\frac{\langle\epsilon\rangle_{V,t}}{u^3_{rms}}\,.
\end{equation}
Figure \ref{c_eps} shows the friction coefficient as a function of the Rayleigh number which can be fitted by a power law with an exponent of -0.16.
We have added the slopes which would follow from a laminar flow, $c_{\epsilon}\sim Re^{-1} \sim Ra^{-1/2}$, and a fully turbulent wall bounded flow,
$c_{\epsilon}\sim Re^{-1/4} \sim Ra^{-1/8}$ (at moderate Reynolds numbers), given the simple $Re$--$Ra$ scaling of $Re\sim \sqrt{Ra}$ which is almost satisfied for our data as mentioned in 
Section \ref{sec:level2}. Our detected slope is also consistent with the first data points of Fig. 16(a) in \cite{Verzicco2003} (denoted as the surrogate friction 
coefficient therein). If we compare to experiments for water, $\Gamma = 1$ and Rayleigh numbers  overlapping with ours, our exponent disagrees with \cite{Sun2008} who measured an exponent of $-0.28$ (their experiment was in an elongated  rectangular box) but agrees with \cite{Wei2013} who measured an exponent of $-0.19\pm 0.02$ (their experiment was in a cylindrical container).

The behavior of the friction coefficient indicates that the boundary layer dynamics, in which the major part of the kinetic energy is dissipated, is a 
mixture of laminar phases interrupted by bursts due to rising plumes and local vortices, i.e., elements which are characteristic of a turbulent boundary 
layer. This has also been found experimentally by \cite{duPuits2014} and also numerically by \cite{Shi2012}, who performed a detailed analysis of the 
dynamics in the vicinity of the heating plate.   

\begin{figure}
\centering
\includegraphics[width=0.8\textwidth]{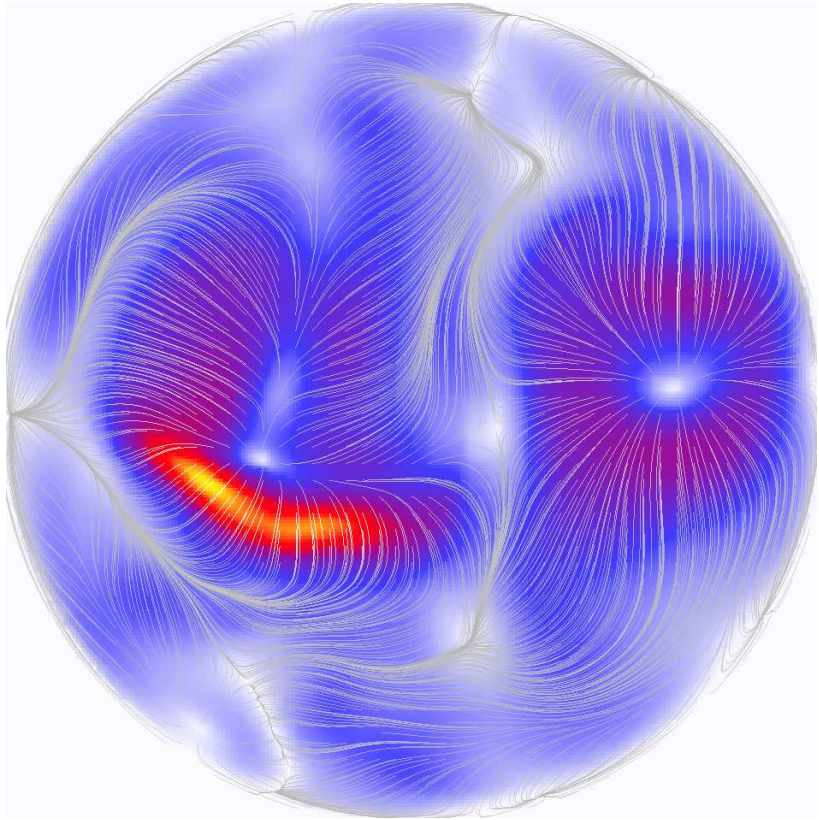}
\caption{Color density plot of the instantaneous magnitude of $\tau_w$ (calculated from the magnitude of (\ref{flux2})) along with the gray field lines of the wall shear stress vector field 
$\left(\frac{du}{dz}, \frac{dv}{dz}\right)$ at the bottom plate for $Ra=1\times 10^7, \Gamma = 1.0$. Color coding is as follows. White is the minimum and yellow denotes 
maximum magnitude.}
\label{tauw1}
\end{figure}

Another perspective is provided by \cite{Chong2012} who found critical points (i.e. nodes/foci) in the wall shear stress vector field in a turbulent channel flow simulations. 
Similar patterns have been reported in \cite{Grosse2009} for an experimental determination of the wall-stress vector field with micro-pillars in a duct flow. 
These critical points were rare, but did exist and they postulated that these critical points would give rise to a generation of turbulence at the wall. We also find such critical points in our simulations as seen in Figure \ref{tauw1} where we have plotted the wall shear stress vector field lines superimposed on a color density 
plot of the magnitude of $\tau_w$ for a Rayleigh number of $1\times 10^7$ and $\Gamma = 1$. The white areas contain exactly those zero points of the vector field
that have been excluded in our velocity boundary layer scale analysis. Their role for the enstrophy production will be discussed in a subsequent work.

\section{Summary and outlook}
This paper presented a boundary layer analysis in turbulent Rayleigh-B\'{e}nard convection. The motivation for this work is threefold.

First, we wish to generalize the boundary layer analysis to a fully local approach. This perspective best captures the spatio-temporal
variations and the transitional character of the boundary layer dynamics in which quasi-laminar sequences are interrupted by bursts and vortex 
formation processes which are partly connected to the detachment of plumes.

Second, the local definition is to our view necessary since the near-wall
boundary layer dynamics is highly inhomogeneous across the bottom and top plates in closed convection cells which are the standard experimental setup. Recall that a well-defined mean flow as in a pipe or channel is absent here. The large-scale circulation itself is a complex three-dimensional time-dependent structure filling the whole cell.
We demonstrate this circumstance for example by 
a comparison of runs at the same $Ra$ and $Pr$, but different aspect ratio. It matters where the data are taken: away from the side walls, 
in the middle of the cell or in between a multi-roll large-scale circulation configuration which can build up in cells with aspect ratios $\Gamma>1$. 
In each of these cases the tails of the local boundary layer scale distributions are found to differ slightly, hence prefactors and scaling exponents in 
power laws for the first moments derived from the distributions will vary. The study also complements previous experimental and numerical local boundary layer analyses.
It confirms the inhomogenenous character of the dynamics and resulting statistics which is found to depend on the location at the plate.

Third, our analysis makes also the first contact to recent efforts to study local dissipation 
scales and higher order statistics of velocity gradients in bulk turbulence of  several turbulent flows. The distribution of the local boundary layer scales 
is thus a direct manifestation of the strong spatial intermittency of the gradients in the boundary layer. 

With this fully local boundary layer analysis, we are able to assemble good statistics on the distribution of boundary layer thicknesses.
We find that the overall shape of the PDFs for both inner and outer and thermal and velocity boundary layer thicknesses take on a universal shape, but one which is not lognormal. The scaling exponents of the mean (first moment) outer boundary layer thicknesses agree well with previous results obtained from the more traditional slope method, when similar plate averagings are compared. However, the mean inner velocity boundary layer thicknesses tend to scale more steeply. We also are able to compute a shear Reynolds number from our data and find that our scaling exponent for the bulk-averaged method ($Re_{s,b} \propto Re^{0.24\pm 0.01}$)  agrees with other results ($Re_s \propto Ra^{0.267\pm0.0386}$). We also computed dissipation layer thicknesses and found a transition in both thickness scalings with Rayleigh number at around $1\times 10^7 < Ra < 1\times 10^8$, exactly where a transition to small-scale turbulence in the bulk is expected from \cite{Schumacher2014}. Finally our friction coefficient scaling with Rayleigh number ($c_{\epsilon} \propto Ra^{-0.16\pm 0.01}$) suggests that the boundary layer is in a transitional regime.

The present study can be a first step only. A formal conclusion on the transitional behavior of the boundary layer requires data at larger Rayleigh 
numbers and larger aspect ratios as well.  A further aspect to study is the Prandtl number dependence. This would also help to better understand
the parameter dependence of the dissipation layers in comparison to the other scales. In the present study the thermal and velocity boundary layers are similar in size since the Prandtl number is close to one, but for Prandtl numbers both larger and smaller than one the boundary layer thicknesses will differ. In particular for very low Prandtl numbers, we can expect dramatic changes in the boundary layer dynamics and the related turbulence production mechanisms based on our current efforts in this 
direction which will be discussed elsewhere.

\begin{acknowledgments}
This work was supported by the Deutsche Forschungsgemeinschaft within Research Unit 1182. Supercomputing time 
for the majority of the present RBC simulations has been provided on BG/Q JUQUEEN by the J\"ulich Supercomputing Centre (Germany) within the 
Large-Scale Project HIL07 of the German Gauss Centre for Supercomputing. The simulation run for the larger-aspect ratio 
cell ($\Gamma=3$) was made possible by the PRACE Initiative. We acknowledge that the results of this research have been achieved
using the PRACE-2IP project (FP7 RI-283493) resource BG/Q Hartree based in UK at Daresbury Laboratory. Both of us wish to thank 
also Paul Fischer for his initial help with the Nek5000 spectral element code package and Bruno Eckhardt for discussions.
\end{acknowledgments}

\end{document}